\documentclass[
superscriptaddress,
onecolumn,
prd,tightenlines,
nofootinbib, eqsecnum,
amsfonts,amsmath,amssymb]{revtex4}

\usepackage{bm}
\usepackage{dsfont}
\usepackage{graphicx} 
\usepackage{hyperref}
\usepackage{mathrsfs}
\usepackage{multirow}

\newcommand{\ud}{\mathrm{d}}

\usepackage{ulem}
\usepackage[usenames]{color}

\allowdisplaybreaks

\begin{document}

\title{Quantum Tests of the Einstein Equivalence Principle with the
  \\STE-QUEST Space Mission}

\author{Brett Altschul}\affiliation{Department of Physics and
  Astronomy, University of South Carolina Columbia, SC 29208, USA}
\author{Quentin G. Bailey}\affiliation{Physics Department,
  Embry-Riddle Aeronautical University, 3700 Willow Creek Road,
  Prescott, Arizona 86301, USA}
\author{Luc Blanchet}\affiliation{GReCO, Institut d'Astrophysique de
  Paris, CNRS UMR 7095, UPMC, 98$^\text{bis}$ boulevard Arago, 75014
  Paris, France}
\author{Kai Bongs}\affiliation{Midlands Ultracold Atom Research
  Centre, School of Physics and Astronomy, University of Birmingham,
  Birminham B15 2TT, UK}
\author{Philippe Bouyer}\affiliation{LP2N, IOGS, CNRS, Universit\'e de
  Bordeaux, Institut d'Optique, avenue Fran\c{c}ois Mitterrand, 33405
  Talence, France}
\author{Luigi Cacciapuoti}\affiliation{European Space Agency,
  Keplerlaan 1 -- P.O. Box 299, 2200 AG Noordwijk ZH, The Netherlands}
\author{Salvatore Capozziello}\affiliation{Dipartimento di Fisica,
  Universit\`a di Napoli ``Federico II'', Complesso Universitario di
  Monte S. Angelo, Via Cinthia, Ed. N I-80126 Napoli,
  Italy}\affiliation{Istituto Nazionale di Fisica Nucleare, Sez. di
  Napoli, Via Cinthia, Ed.  N I-80126 Napoli, Italy}\affiliation{Gran
  Sasso Science Insitute (INFN), Viale F. Crispi 7, I-67100, L'Aquila,
  Italy}
\author{Naceur Gaaloul}\affiliation{Institut f\"ur Quantenoptik,
  Leibniz Universit\"at Hannover, Welfengarten 1, 30167 Hannover,
  Germany}
\author{Domenico Giulini}\affiliation{Center of Applied Space
  Technology and Microgravity, University of Bremen, Am Fallturm 1,
  D-28359 Bremen, Germany}\affiliation{Institute for Theoretical
  Physics, Leibniz Universit\"at Hannover, Appelstrasse 2, D-30167
  Hannover, Germany}
\author{Jonas Hartwig}\affiliation{Institut f\"ur Quantenoptik,
  Leibniz Universit\"at Hannover, Welfengarten 1, 30167 Hannover,
  Germany}
\author{Luciano Iess}\affiliation{Dipartimento di Ingegneria Meccanica
  e Aerospaziale, Sapienza Universit\`a di Roma, via Eudossiana 18,
  00184, Rome, Italy}
\author{Philippe Jetzer}\affiliation{Physik-Institut, Universit\"at
  Z\"urich, Winterthurerstrasse 190, CH-8057 Z\"urich, Switzerland}
\author{Arnaud Landragin}\affiliation{SYRTE, CNRS, Observatoire de
  Paris, LNE, UPMC, 61 avenue de l'Observatoire, 75014 Paris, France}
\author{Ernst Rasel}\affiliation{Institut f\"ur Quantenoptik, Leibniz
  Universit\"at Hannover, Welfengarten 1, 30167 Hannover, Germany}
\author{Serge Reynaud}\affiliation{Laboratoire Kastler Brossel, CNRS,
  ENS, UPMC, Campus Jussieu, F-75252 Paris, France}
\author{Stephan Schiller}\affiliation{Institut f\"ur
  Experimentalphysik, Heinrich-Heine-Universit\"at D\"usseldorf,
  Universit\"atstrasse 1, 40225 D\"usseldorf, Germany}
\author{Christian Schubert}\affiliation{Institut f\"ur Quantenoptik,
  Leibniz Universit\"at Hannover, Welfengarten 1, 30167 Hannover,
  Germany}
\author{Fiodor Sorrentino}\affiliation{Dipartimento di Fisica e
  Astronomia and LENS, Universit\`a di Firenze - INFN Sezione di
  Firenze, Via Sansone 1, 50019 Sesto Fiorentino, Italy}
\author{Uwe Sterr}\affiliation{Physikalisch-Technische Bundesanstalt
  (PTB), Bundesallee 100, D-38116 Braunschweig, Germany}
\author{Jay D. Tasson}\affiliation{Physics and Astronomy Department,
  Carleton College, One North College Street, Northfield, Minnesota
  55057, USA}
\author{Guglielmo M. Tino}\affiliation{Dipartimento di Fisica e
  Astronomia and LENS, Universit\`a di Firenze - INFN Sezione di
  Firenze, Via Sansone 1, 50019 Sesto Fiorentino, Italy}
\author{Philip Tuckey}\affiliation{SYRTE, CNRS, Observatoire de Paris,
  LNE, UPMC, 61 avenue de l'Observatoire, 75014 Paris, France}
\author{Peter Wolf}\affiliation{SYRTE, CNRS, Observatoire de Paris,
  LNE, UPMC, 61 avenue de l'Observatoire, 75014 Paris, France}

\date{\today}

\begin{abstract}
  We present in detail the scientific objectives in fundamental
  physics of the Space-Time Explorer and QUantum Equivalence Space
  Test (STE-QUEST) space mission. STE-QUEST was pre-selected by the
  European Space Agency together with four other missions for the
  cosmic vision M3 launch opportunity planned around 2024. It carries
  out tests of different aspects of the Einstein Equivalence Principle
  using atomic clocks, matter wave interferometry and long distance
  time/frequency links, providing fascinating science at the interface
  between quantum mechanics and gravitation that cannot be achieved,
  at that level of precision, in ground experiments. We especially
  emphasize the specific strong interest of performing equivalence
  principle tests in the quantum regime, \textit{i.e.} using quantum
  atomic wave interferometry. Although STE-QUEST was finally not
  selected in early 2014 because of budgetary and technological
  reasons, its science case was very highly rated. Our aim is to
  expose that science to a large audience in order to allow future
  projects and proposals to take advantage of the STE-QUEST
  experience.
\end{abstract}


\maketitle


\section{Introduction}
\label{sec:intro}

\subsection{Scientific Motivations}
\label{sec:scmotiv}

Our best knowledge of the physical Universe, at the deepest
fundamental level, is based on two theories: Quantum Mechanics (or,
more precisely, Quantum Field Theory) and the classical theory of
General Relativity. Quantum Field Theory has been extremely successful
in providing an understanding of the observed phenomena of atomic,
particle, and high energy physics and has allowed a unified
description of three of the four fundamental interactions that are
known to us: electromagnetic, weak and strong interactions (the fourth
one being gravitation). It has led to the Standard Model of particle
physics that has been highly successful in interpreting all observed
particle phenomena, and has been strongly confirmed with the recent
discovery at the LHC of the Higgs (or, more precisely,
Brout-Englert-Higgs) boson, which could in fact be viewed as the
discovery of a fifth fundamental interaction. Although open questions
remain within the Standard Model of particle physics, it is clearly
the most compelling model for fundamental interactions at the
microscopic level that we have at present.

On the other hand, Einstein's theory of General Relativity (GR) is a
cornerstone of our current description of the physical world at
macroscopic scales. It is used to understand the flow of time in the
presence of gravity, the motion of bodies from satellites to galaxy
clusters, the propagation of electromagnetic waves in the vicinity of
massive bodies, the evolution of stars, and the dynamics of the
Universe as a whole. GR brilliantly accounts for all observed
phenomena related to gravitation, in particular all observations in
the Earth's environment, the Solar system, in relativistic binary
pulsars and, beyond that, on galactic and cosmological scales.

The assumed validity of GR at cosmological scales, and the fact that
non-gravitational interactions are described by the Standard Model of
particle physics, together with a hypothesis of homogeneity and
isotropy of cosmological solutions of these theories, have led to the
``concordance model'' of cosmology, referred to as the $\Lambda$-CDM
(Cold Dark Matter) model, which is in agreement with all present-day
observations at large scales, notably the most recent observations of
the anisotropies of the cosmic microwave background by the Planck
satellite~\cite{Ade2013}. However, important puzzles remain, in
particular the necessary introduction of dark energy, described by a
cosmological constant $\Lambda$, and of cold dark matter, made of some
unknown, yet to be discovered, stable particle.

There is a potential conflict on the problem of dark matter between
the concordance model of cosmology and the Standard Model of
particles. On the one hand, there is strong evidence~\cite{Ade2013}
that 26.8 \% of the mass-energy of the Universe is made of
non-baryonic dark matter particles, which should certainly be
predicted by some extension of the Standard Model of particles. On the
other hand, there is no indication of new physics beyond the Standard
Model which has been found at the LHC. For instance, the search of
supersymmetry at LHC has for the moment failed.

Although very successful so far, GR as well as numerous other
alternative or more general theories of gravitation are classical
theories. As such, they are fundamentally incomplete, because they do
not include quantum effects. A theory solving this problem would
represent a crucial step towards the unification of all fundamental
forces of Nature. Most physicists believe that GR and the Standard
Model of particle physics are only low-energy approximations of a more
fundamental theory that remains to be discovered. Several concepts
have been proposed and are currently under investigation
(\textit{e.g.}, string theory, loop quantum gravity, extra spatial
dimensions) to bridge this gap and most of them lead to tiny
violations of the basic principles of GR.

One of the most desirable attributes of that fundamental theory is the
unification of the fundamental interactions of Nature, \textit{i.e.} a
unified description of gravity and the three other fundamental
interactions. There are several attempts at formulating such a theory,
but none of them is widely accepted and considered
successful. Furthermore, they make very few precise quantitative
predictions that could be verified experimentally. One of them is the
Hawking radiation of black holes, which is however far from being
testable experimentally for stellar-size black holes we observe in
astrophysics.

Therefore, a fuller understanding of gravity will require observations
or experiments able to determine the relationship of gravity with the
quantum world. This topic is a prominent field of activity with
repercussions covering the complete range of physical phenomena, from
particle and nuclear physics to galaxies and the Universe as a whole,
including dark matter and dark energy.

A central point in this field is that most unification theories have
in common a violation at some (\textit{a priori} unknown) level of one
of the basic postulates of GR, which can be tested experimentally: the
Einstein Equivalence Principle (EEP). Let us emphasize that the Weak
Equivalence Principle (WEP) is not a fundamental symmetry of physics,
contrary to \textit{e.g.} the principle of local gauge invariance in
particle physics. An important challenge is therefore to test with the
best possible accuracy the EEP. This is then the main motivation of
many experiments in fundamental physics, both on Earth and in space.

Precision measurements are at the heart of the scientific method that,
since Galileo's time, is being used for unveiling Nature and
understanding its fundamental laws. The assumptions and predictions of
GR can be challenged by precision experiments on scales ranging from
micrometers in the laboratory to the Solar System size, in the latter
case using spacecrafts or the orbiting Earth, Moon and planets. The
implementation of tests with significantly improved sensitivity
obviously requires the use of state-of-the-art technology, and in case
of satellite-based experiments the challenge is to make such
technology compatible with use in space, \textit{i.e.} extremely
robust, reliable, and automatized.

\subsection{The Space Mission STE-QUEST}
\label{sec:mission}

The satellite STE-QUEST (Space-Time Explorer and QUantum Equivalence
Space Test) is specifically designed for testing different aspects of
the EEP and searching for its violation with high precision. The weak
equivalence principle has been verified with high precision using
torsion balances on ground~\cite{Schlamminger2008} and the Lunar laser
ranging~\cite{Williams2004}. It will be tested in Earth orbit by the
CNES satellite $\mu$-SCOPE (Micro-Satellite \`a tra\^in\'ee
Compens\'ee pour l'Observation du Principe d'Equivalence) in
2016~\cite{Touboul2001}. On the other hand, the gravitational
red-shift, a different aspect of the EEP, was first measured using
gamma ray spectroscopy in the laboratory~\cite{Pound1960}, and the
most precise test so far was done in space with the GP-A
experiment~\cite{Vessot1979}. The ESA mission ACES (Atomic Clock
Ensemble in Space) will test the gravitational red-shift with the
highly accurate laser-cooled atomic clock PHARAO on the International
Space Station (ISS) in 2016~\cite{Cacciapuoti2009}.

Atomic clocks and high-performance time and frequency links, atom
interferometers and classical accelerometers are today able to measure
frequency, time, and distances, and furthermore to track the motion of
massive bodies, quantum particles, and light to accuracy levels never
reached before. These instruments achieve their ultimate performance
in space, where the clean environment and the free-fall conditions
become essential for identifying tiny deformations in space-time that
might bring the signature of new physics or new fundamental
constituents. From this point of view, it is not surprising that
fundamental physics pervades all aspects of space science.

STE-QUEST was proposed in the fall of 2010 in response to ESA's M3
call in the Cosmic Vision programme (with launch date in the 2022-24
time interval), by a science team under coordination by S.~Schiller
and E.M.~Rasel with support from 67 colleagues from Europe and the
USA. STE-QUEST is based on the earlier proposals
EGE~\cite{Schiller2009} and MWXG~\cite{Ertmer2009}, submitted to ESA's
M2 call. ESA performed a ``concurrent design facility'' study of a
mission concept similar to EGE, named STE, in 2010. Previously, ESA
had also convened a Fundamental Physics Advisory Team which in 2009-10
developed a roadmap on fundamental physics in space.

STE-QUEST, together with three other mission proposals, was selected
in early 2011 by ESA's advisory structure as one candidate
mission. STE-QUEST went through an assessment phase study of the
satellite and payload (see the Yellow Book of the
mission~\cite{YellowBook}). As a result of the assessment phase and in
agreement with the national space agencies, STE-QUEST was removed from
the candidate pool in late 2013, before the final selection of a
single mission for the M3 slot, because of budgetary and technological
reasons. Nevertheless, ESA's advisory committees evaluated the science
aspects of STE-QUEST in early 2014 (together with the remaining M3
candidates) and ranked them highly. It is likely that STE-QUEST will
recompete for the M4 launch slot.

The primary science objectives of STE-QUEST is testing the different
aspects of the Einstein Equivalence Principle with quantum
sensors. The payload consists of a differential atom interferometer
comparing the free propagation of matter waves of different
composition under the effect of gravity and a frequency comparison
link in the microwave domain for comparing atomic clocks on
ground. STE-QUEST performs a direct test of the WEP by comparing the
free fall of quantum objects of different composition. The E\"otv\"os
ratio between the matter waves of two isotopes of the Rubidium atom is
measured in a differential atom interferometer down to the $2 \times
10^{-15}$ uncertainty level. While present limits on WEP tests
involving classical objects reach an uncertainty of a few parts in
$10^{13}$, measurements performed on quantum objects (matter waves in
states which have no classical counterpart, \textit{e.g.}
spatio-temporal quantum superpositions) are still at the level of a
few parts in $10^7$~\cite{Fray2004, Schlippert2014, Tarallo2014}. From
this point of view, STE-QUEST will explore the boundaries between
gravitation and quantum mechanics, significantly improving existing
measurements and complementing experiments such as $\mu$-SCOPE,
designed for a classical WEP test in space to the level $1 \times
10^{-15}$.
\begin{figure}[t]
\begin{center}
\begin{tabular}{c}
\includegraphics[width=14cm]{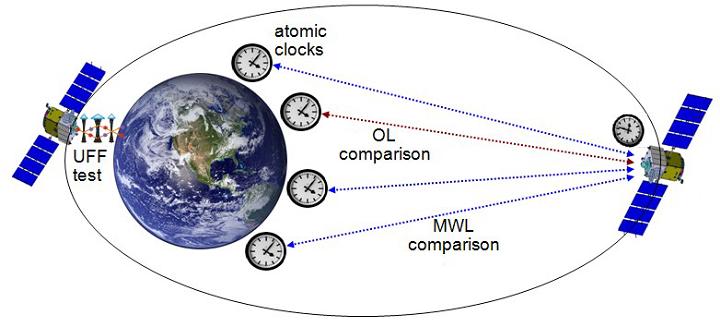}
	\end{tabular}
        \caption{The STE-QUEST spacecraft in orbit around the
          Earth. The mission is designed to test the Einstein
          Equivalence Principle by tracking the free-fall motion of
          quantum matter waves, by performing gravitational red-shift
          tests between ground clocks on intercontinental distances
          and with (optionally) a high-stability and high-accuracy
          onboard clock, and by performing tests of local Lorentz
          invariance and CPT symmetry. (OL: optical link; MWL:
          microwave link.)}\label{fig:stequest}
\end{center}
\end{figure}
\begin{table*}[]
\begin{center}
  \begin{tabular}{|p{4.7cm}||p{11.3cm}|}
    \hline \centerline{\textbf{Science Investigation}} &
    \centerline{\textbf{Measurement Requirement}} \\\hline
    \multicolumn{2}{|l|}{\textbf{Weak Equivalence Principle Tests}}
    \\ \hline \textit{Universality of propagation of matter-waves} &
    Test the universality of the free propagation of matter waves to an
    uncertainty in the E\"otv\"os parameter better than $2\times
    10^{-15}$. \\ \hline \multicolumn{2}{|l|}{\textbf{Gravitational
        Red-shift Tests}}\\ \hline \textit{Sun gravitational red-shift}
    & Test of the Sun gravitational red-shift effect to a
    fractional frequency uncertainty of $2\times 10^{-6}$, with an
    ultimate goal of $5\times 10^{-7}$. \\ \hline \textit{Moon
      gravitational red-shift} & Test of the Moon gravitational
    red-shift effect to a fractional frequency uncertainty of $4\times
    10^{-4}$, with an ultimate goal of $9\times 10^{-5}$. \\ \hline
    \textit{Earth gravitational red-shift (optional)\footnote{This
        scientific investigation can be performed only if the STE-QUEST
        payload is equipped with a high-stability and high-accuracy
        atomic clock.}} & Measurement of the Earth gravitational
    red-shift effect to a fractional frequency uncertainty of $2\times
    10^{-7}$. \\ \hline \multicolumn{2}{|l|}{\textbf{Local Lorentz
        Invariance and CPT Tests}}\\ \hline \textit{LLI and CPT} &
    Provide significant improvements on the determination of several LLI
    and CPT parameters of the Lorentz and CPT symmetry violating
    Standard Model Extension. \\ \hline
\end{tabular}
\caption{Science investigations \textit{vs.} measurement requirements
  for topics in fundamental physics that shall be investigated by
  STE-QUEST.}\label{tab:topics}
\end{center}
\end{table*}

STE-QUEST also tests another complementary aspect of the Einstein
Equivalence Principle, one of the most fascinating effects predicted
by GR and other metric theories of gravity: the gravitational
red-shift or gravitational time dilation effect. As direct consequence
of the EEP, time runs (or clocks tick) more slowly near a massive
body. This effect can be detected when comparing the time intervals
measured by identical clocks placed at different depths in a
gravitational field. The microwave link (MWL) of the STE-QUEST
satellite allows comparing ground clocks down to the $1 \times
10^{-18}$ uncertainty level. Such measurements, far beyond the
capabilities of existing frequency transfer systems, will perform
clock red-shift tests in the field of the Sun and the Moon,
respectively at the $2 \times 10^{-6}$ and $4 \times 10^{-4}$
uncertainty levels. For comparison, existing measurements of the Sun
red-shift effect are at the few \% uncertainty level while, to our
knowledge, no such tests have ever been performed in the field of the
Moon. An optional (depending on available funding) onboard clock
allows additionally a red-shift measurement in the Earth field by
taking advantage of the high apogee and high eccentricity of the
orbit. The clock under consideration is derived from the PHARAO cold
atom \text{Cs} clock to be flown on the
ISS~\cite{Cacciapuoti2009}. The version planned for STE-QUEST is
designed to reach an uncertainty in the Earth field red-shift test of
$2 \times 10^{-7}$, one order of magnitude better than the objective
of ACES. The relativistic theory for time and frequency transfer
needed for frequency links in space missions such as ACES and
STE-QUEST is described in Ref.~\cite{Blanchet2001}.

Clock red-shift measurements obtained in the field of the Earth, the
Sun or the Moon test the Local Position Invariance (LPI) principle and
search for anomalous couplings depending on the composition of the
source of the gravitational field. LPI is a constituent of EEP
together with WEP and the Local Lorentz Invariance (LLI) principle,
see Sec.~\ref{sec:facet}. As we shall discuss in
Sec.~\ref{sec:diffEEP}, in generic frameworks modelling a possible
violation of EEP, WEP and clock red-shift tests are complementary and
need to be pursued with equal vigor as, depending on the model used,
either one of the tests can prove significantly more sensitive than
the other. Improving the accuracy of these tests will bring
significant progress in restricting the parameters space and
discriminating between theories seeking to unify quantum mechanics
with gravity. The eventual detection of an EEP violation would carry
the signature of new fundamental constituents or interactions in the
Universe (\textit{e.g.} scalar fields for dark energy, particles for
dark matter, fundamental strings, \textit{etc.}). In this case,
STE-QUEST tests would have a significant impact not only for
fundamental physics research, but also for cosmology and particle
physics. The ensemble of fundamental physics science objectives of
STE-QUEST is summarized in Table~\ref{tab:topics} and
Fig.~\ref{fig:stequest}.

STE-QUEST has also important applications in domains other than
fundamental physics, in particular in the fields of time and frequency
metrology and for geodesy studies. As mentionned, the STE-QUEST
high-performance MWL provides the means for connecting atomic clocks
on ground in a global network, enabling comparisons down to the $1
\times 10^{-18}$ fractional frequency uncertainty level. Clock
comparisons \textit{via} STE-QUEST will contribute to the realization
of international atomic time scales (UTC, TAI, \textit{etc.}) and to
the improvement of their stability and accuracy. Synchronization of
clocks, space-to-ground and ground-to-ground, to better than
$50\,\text{ps}$ can be achieved through STE-QUEST for distributing
time scales to unprecedented performance levels. Common-view
comparisons of ground clocks, primarily used for gravitational
red-shift tests in the field of Sun or Moon, also provide direct
information on the geopotential differences at the locations of the
two ground clocks. STE-QUEST will therefore contribute to establishing
a global reference frame for the Earth gravitational potential at the
sub-$\text{cm}$ level through local measurements. This method is
complementary to current and future satellite gravimetry missions such
as CHAMP, GRACE and GOCE as well as to altimetry missions like JASON
and Envisat in defining the Global Geodetic Observing System
(GGOS). The Table~\ref{tab:othertopics} (relegated in the conclusion
section~\ref{sec:conclusion}) summarizes the list of topics other than
fundamental physics that shall be investigated by STE-QUEST.

The present paper is an adapted version of the fundamental physics
science objectives of STE-QUEST extracted from the Yellow Book of
STE-QUEST which is available in Ref.~\cite{YellowBook} (see also
Ref.~\cite{Aguilera2013}). The Yellow Book also gives an overview of
science objectives in other fields (geodesy, time/frequency metrology,
reference frames) and details on the mission and payload, which are
however beyond the scope of this paper that focuses on the fundamental
physics objectives. In Sec.~\ref{sec:EEP} we shall review in more
detail the EEP and its different facets. In Sec.~\ref{sec:EEPtoday} we
shall discuss the status of EEP in Physics today and particularly in
the contexts of cosmology and particle physics. Quantum mechanics and
the EEP and the potential interest of quantum tests of the EEP will be
analyzed in Sec.~\ref{sec:quantum}. The specific tests of the EEP
which will be achieved by STE-QUEST will be presented in
Sec.~\ref{sec:STEtest}. The paper ends with the main conclusions in
Sec.~\ref{sec:conclusion}.

\section{The Einstein Equivalence Principle}
\label{sec:EEP}

\subsection{The Different Aspects of the EEP}
\label{sec:facet}

The foundations of gravitational theories and the equivalence
principle have been clarified by many authors, including
Schiff~\cite{Schiff1960}, Dicke~\cite{Dicke1964}, Thorne, Lee \&
Lightman~\cite{Thorne1973}, and others. Following the book of
Will~\cite{Will1993} the EEP is generally divided into three
sub-principles: the Weak Equivalence Principle (WEP) also known as the
Universality of Free Fall (UFF), Local Lorentz Invariance (LLI), and
Local Position Invariance (LPI). The EEP is satisfied if and only if
all three sub-principles are satisfied. Below we describe these three
sub-principles:
\begin{enumerate}
\item WEP (or UFF) states that if any uncharged test body\footnote{By
  test body is meant an electrically neutral body whose size is small
  enough that the coupling to inhomogeneities in the gravitational
  field can be neglected.} is placed at an initial event in space-time
  and given an initial velocity there, then its subsequent trajectory
  will be independent of its internal structure and composition. The
  most common test of WEP consists in measuring the relative
  acceleration of two test bodies of different internal structure and
  composition freely falling in the same gravitational field. If WEP
  is satisfied, that relative acceleration is zero;
\item LLI states that the outcome of any local non-gravitational test
  experiment is independent of the velocity and orientation of the
  (freely falling) apparatus. Tests of LLI usually involve a local
  experiment (\textit{e.g.} the comparison of the frequency of two
  different types of clocks) whose velocity and/or orientation is
  varied in space-time. LLI is verified if the result of the
  experiment is unaltered by that variation;
\item LPI states that the outcome of any local non-gravitational test
  experiment is independent of where and when in the Universe it is
  performed. Tests of LPI usually involve a local experiment
  (\textit{e.g.} the measurement of a fundamental constant, or the
  comparison of two clocks based on different physical processes) at
  different locations and/or times. In particular, varying the local
  gravitational potential allows for searches of some anomalous
  coupling between gravity and the fields involved in the local
  experiment. A particular version of such tests, known as test of the
  gravitational red-shift, uses the same type of clock, but at two
  different locations (different local gravitational potentials) and
  compares them \textit{via} an electromagnetic signal. Then it can be
  shown (see Sec.~2.4c in Ref.~\cite{Will1993}) that the measured
  relative frequency difference is equal to $\Delta U/c^2$ (where
  $\Delta U$ is the difference in gravitational potential) if and only
  if LPI is satisfied.
\end{enumerate}
One of the unique strengths of STE-QUEST is that it will test all
three aspects of the EEP, using a combination of measurements in space
and on the ground (relative acceleration of different atomic isotopes,
comparison of distant clocks). Additionally, the explored domain of
the possible violation of the LLI and LPI is maximized by the large
variation of velocity and gravitational potential using a highly
elliptic orbit of the spacecraft.

Since the three sub-principles described above are very different in
their empirical consequences, it is tempting to regard them as
independent. However, it was realized quite early that any
self-consistent gravitational theory is very likely to contain
connections between the three sub-principles. This has become known as
Schiff's conjecture~\cite{Schiff1960}, formulated around 1960. Loosely
stated, the Schiff conjecture implies that if one of the three
sub-principles is violated, then so are the other two. This conjecture
can be understood heuristically by the following example. Suppose that
WEP/UFF is violated; then two different clocks (with different
internal compositions) will acquire different accelerations in a
gravitational field. In the freely falling frame of one of the clocks,
the other one will be accelerated (even though being located at the
same position), and there will be an abnormal red-shift between the
two clocks depending on their difference of internal composition,
hence a violation of LPI. The Schiff conjecture has been proved within
very general theoretical frameworks such as the Lagrangian formalism
we shall review in Sec.~\ref{sec:diffEEP}. Alternative theories which
do not satisfy the conjecture suffer from serious pathologies and are
non-viable.

Schiff's conjecture has given rise to much debate, in particular
concerning its empirical consequences and the relative merit of tests
of the different sub-principles. Whilst it is true that any theory
respecting energy conservation (\textit{e.g.} based on an invariant
action principle) must satisfy Schiff's conjecture, the actual
quantitative relationship between violation of the sub-principles is
model dependent and varies as a function of the mechanism used for the
violation (see \textit{e.g.} Sec.~\ref{sec:diffEEP} for a
phenomenological example). As a consequence, it is not known \textit{a
  priori} which test (WEP/UFF, LLI, or LPI) is more likely to first
detect a violation and the most reasonable approach is to perform the
tests of the three sub-principles. This is the philosophy of
STE-QUEST.

For completeness, and to avoid possible confusion, we will say a few
words about the Strong Equivalence Principle (SEP), although it is not
directly related to, and will not be tested by STE-QUEST. The SEP is a
generalization of EEP to include ``test'' bodies with non-negligible
self-gravitation, together with experiments involving gravitational
forces (\textit{e.g.} Cavendish-type experiments). Obviously, SEP
includes EEP as a special case in which gravitational forces can be
ignored. Typical tests of SEP involve moons, planets, stars or local
gravitational experiments, the best known example being lunar laser
ranging that tests the universality of free fall, with the two test
bodies being the Moon and the Earth falling in the field of the
Sun. Clearly the two test bodies have non-negligible self-gravitation
and thus provide a test of SEP. The empirical consequences of SEP and
EEP are quite different; in general a violation of SEP does not
necessarily imply a violation of EEP. Similarly the theoretical
consequences are very different: a violation of EEP excludes not only
GR as a possible theory of gravitation, but also all other metric
theories (\textit{e.g.} all PPN theories, Brans-Dicke theory,
\textit{etc.}). A violation of SEP on the other hand excludes GR, but
allows for a host of other metric theories (\textit{e.g.} PPN theories
that satisfy a particular combination of PPN parameters). In that
sense, SEP and EEP tests are complementary and should be carried out
in parallel within experimental and observational
possibilities. STE-QUEST focuses on EEP, but this does not preclude
the interest of SEP tests like continued and improved lunar laser
ranging.

\subsection{The Role of EEP in Theories of Gravitation}
\label{sec:roleEEP}

The EEP is the foundation of all curved space-time or ``metric''
theories of gravitation, including of course GR. It divides
gravitational theories in two classes: metric theories, those that
embody EEP and non-metric theories, those that do not. This
distinction is fundamental, as metric theories describe gravitation as
a geometric phenomenon, namely an effect of curvature of space-time
itself rather than a field over space-time, quite unlike any of the
other known interactions. It might thus appear unnatural to use a
metric theory for gravitation, so different from the formalisms of the
other interactions, and indeed most unification attempts cast doubt on
precisely this hypothesis and thus on the validity of the EEP. Only
experimental tests can settle the question and, in the light of the
above, experimentally testing the EEP becomes truly fundamental.  To
be more precise (see \textit{e.g.} Refs.~\cite{Dicke1964, Thorne1973,
  Will1993}), a metric theory of gravitation is one that satisfies the
following postulates:
\begin{enumerate}
\item Space-time is endowed with a metric tensor $g_{\mu\nu}$, central
  to the metric equation that defines the infinitesimal line element,
  \textit{i.e.} the space-time separation between two events
\begin{equation}\label{metric}
\ud s^2 = g_{\mu\nu}(x^\rho)\ud x^\mu \ud x^\nu\,,
\end{equation}
in some 4-dimensional space-time coordinate system $x^\rho$;
\item The trajectories of freely falling test bodies are geodesics of
  extremal length,
\begin{equation}\label{interval}
\delta \int \ud s = 0\,,
\end{equation}
\textit{i.e.} they depend only on the geometry of space-time, but are
independent of the test body composition;
\item Clocks measure proper time $\tau$ along their trajectory, given
  by
\begin{equation}\label{propertime}
\ud \tau^2 = - \frac{1}{c^2}\ud s^2\,,
\end{equation}
independent of the type of clock used;
\item In local freely falling reference frames, the non-gravitational
  laws of physics (\textit{i.e.} the other three fundamental
  interactions) satisfy the principles of special relativity.
\end{enumerate}
Obviously the above postulates are a direct consequence of the EEP,
for example LLI and LPI are the foundations of points 3 and 4 and WEP
is the basis of point 2. It is important to note that GR is not the
only possible metric theory that satisfies the above
postulates. Indeed, there exist a large number of such theories like
the scalar-tensor Jordan-Brans-Dicke theories~\cite{Jordan1946,
  Brans1961} and their generalizations. These theories differ from GR
in the way that the metric tensor is related to the distribution of
mass-energy through the existence of other fields associated with
gravity (scalar field, vector field, \textit{etc.}).

Theories in which varying non-gravitational coupling constants are
associated with dynamical fields that couple to matter directly are
not metric theories. In such theories, the fine structure constant
$\alpha$ for instance would vary with space and time. Neither, in this
narrow sense, are theories in which one introduces additional fields
(dilatons, moduli) that couple differently to different types of
mass-energy, \textit{e.g.} some versions of Superstring theory. The
fundamental ingredient of all such non-metric theories is
non-universal coupling to gravity of all non-gravitational fields,
\textit{i.e.} the fields of the Standard Model of particle physics. In
metric theories, coupling to the gravitational field is universal, and
as a consequence the metric of space-time can be studied by a variety
of devices made up of different non-gravitational fields and
particles, and, because of universality, the results will be
independent of the device. For instance, the proper time between two
events is a characteristic of space-time and of the location of the
events, not of the clocks used to measure it~\cite{Will1993}.

Thus experimental tests of the EEP are often viewed as tests of the
universal coupling of gravity (through the metric of space-time
$g_{\mu\nu}$) to all non-gravitational fields of the Standard Model of
particle physics~\cite{Damour2008}. Violations occur when the coupling
is dependent on some attribute of the non-gravitational fields at hand
that may be different for different test bodies, \textit{e.g.}
electromagnetic charge, nuclear charge, total spin, nuclear spin,
quark flavor, lepton number, \textit{etc}. Exploring all possibilities
of such anomalous couplings is the fundamental aim of experimental
tests of the EEP. Note also that in any particular experimental
situation, symmetry requires that such anomalous couplings be not only
a function of the composition of the test body, but also of the mass
which is the source of the gravitational field. As a consequence, the
widest possible range of source and test body configurations needs to
be explored when testing the different aspects of EEP, and this is one
of the aims of STE-QUEST, which will test for EEP violation in the
gravitational fields of the Sun and the Moon. Furthermore, although
not discussed further here, the STE-QUEST data can also be analyzed to
search for violation of EEP in other source fields, \textit{e.g.} that
of galactic dark matter as in Ref.~\cite{Schlamminger2008}. Such
future searches will be part of the legacy of STE-QUEST.

\subsection{Why Would the EEP be Violated?}
\label{sec:whyEEP}

\begin{figure}[t]
\begin{center}
\begin{tabular}{c}
\includegraphics[width=14cm]{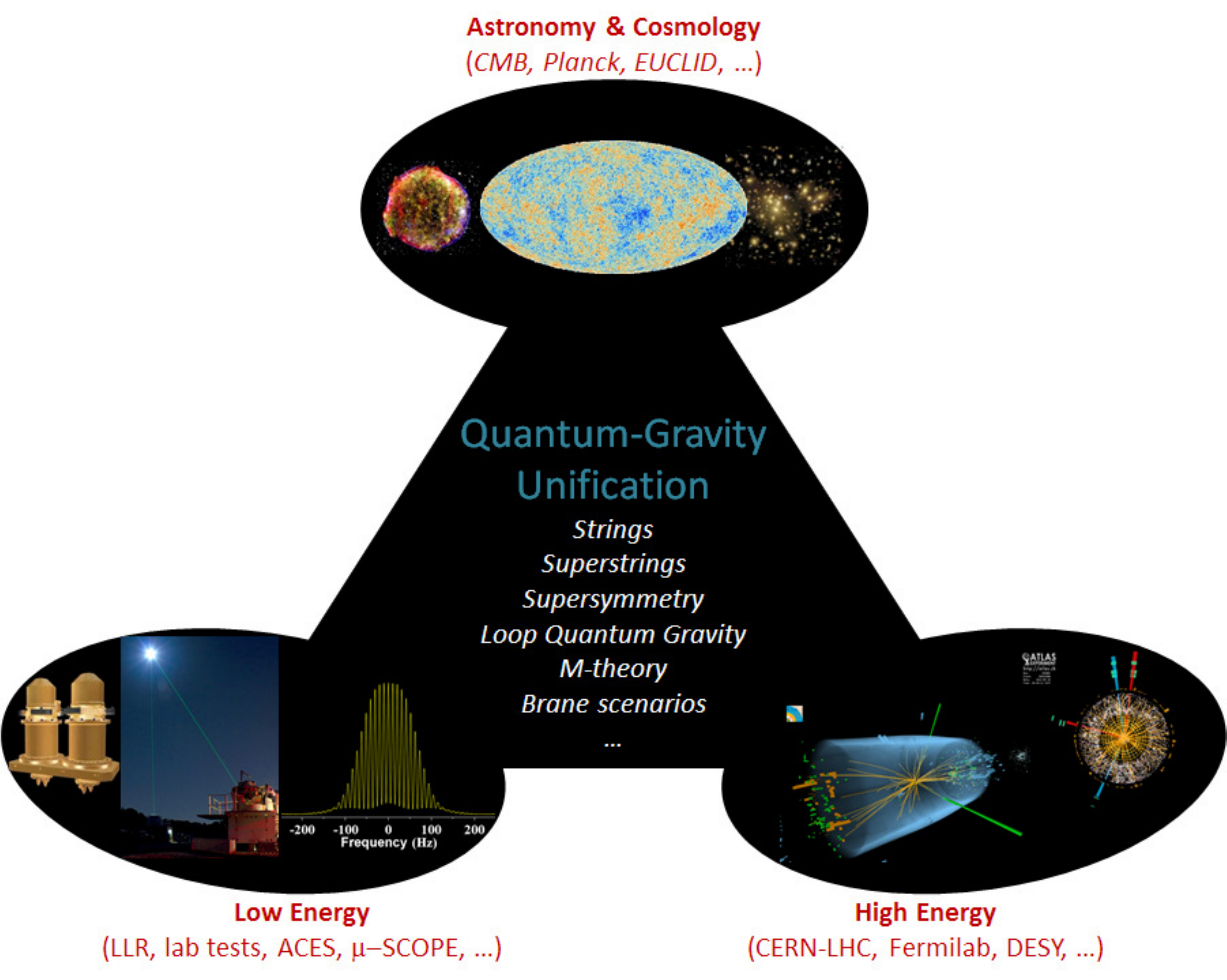}
	\end{tabular}
\caption{Experimental support for quantum gravity and unification
  theories. The relation to cosmology and high energy physics is
  discussed in Sec.~\ref{sec:cosmo} and~\ref{sec:part}. STE-QUEST will
  contribute to the low energy data by improving on several aspects of
  Einstein Equivalence Principle (EEP) tests.}\label{fig:triangleQG}
\end{center}
\end{figure}

It has already been pointed out that the EEP is in fact rather
unnatural in the sense that it renders gravity so different from other
interactions, because the corresponding universal coupling implies
that gravitation is a geometrical attribute of space-time itself
rather than a field over space-time like all other known
interactions. Einstein himself initially called it the
\textit{hypothesis of equivalence} before elevating it to a
\textit{principle} once it became clear how central it was in the
generalization of special relativity to include gravitation. This
shows how surprising it is in fact that such an hypothesis should be
satisfied at all, let alone down to the uncertainties of present-day
tests. Therefore, rather than asking why the EEP should be violated,
the more natural question to ask is why no violation has been observed
yet. Indeed most attempts at quantum gravity and unification theories
lead to a violation of the EEP~\cite{Taylor1988, Damour1994,
  Dimopoulos1996, Antoniadis1998, Rubakov2001, Maartens2010}, which in
general have to be handled by some tuning mechanism in order to make
the theory compatible with existing limits on EEP violation. For
example, in string theory moduli fields need to be rendered massive
(short range)~\cite{Taylor1988} or stabilized by \textit{e.g.}
cosmological considerations~\cite{Damour1994} in order to avoid the
stringent limits already imposed by EEP tests. Similarly M-theory and
Brane-world scenarios using large or compactified extra dimensions
need some mechanism to avoid existing experimental limits from EEP
tests or tests of the inverse square law~\cite{Antoniadis1998,
  Maartens2010, Rubakov2001, Adelberger2009, Antoniadis2011}. The
latter tests explore a modification of the gravitational inverse
square law (\textit{e.g.} in the form of a Yukawa potential) and are
in many respects complementary to EEP tests. However, violations of
the inverse square law will also be detected by certain EEP tests
(\textit{e.g.} red-shift tests), allowing for a much richer
phenomenology with different distance dependences and anomalous
couplings. Therefore, not only do we expect a violation of EEP at some
level, but the non-observation of such a violation with improving
uncertainty is already one of the major experimental constraints for
the development of new theories in the quest for quantum gravity and
unification (see Fig.~\ref{fig:triangleQG}). This makes experimental
tests of EEP in all its aspects one of the most essential enterprises
of fundamental physics today.

It is interesting to note that experimental constraints for EEP
violations at low energy are rather closely related to present-day
physics at the very small scale (particle physics) and the very large
scale (cosmology). These connections are discussed in more detail in
Sec.~\ref{sec:cosmo} and~\ref{sec:part}. Notably, the recent
experimental confirmation of the Higgs boson has thus lent strong
credibility to the existence of scalar fields, as the Higgs is the
first fundamental scalar field observed in Nature. It is thus likely
that additional long and/or short range scalar fields exist, as
postulated by many unification theories, and EEP tests are one of the
most promising experimental means for their observation.

At the other extreme, in cosmology, most models for Dark Energy (DE)
are also based on long-range scalar fields that, when considered in
the context of particle physics, are non-universally coupled to the
fields of the Standard Model~\cite{Khoury2004a, Khoury2004b}. As a
consequence, one would expect EEP violations from such fields at some
level, which might be detectable by experiments like STE-QUEST thus
shedding light on the dark energy content of the Universe from a
completely different angle. Similarly, long-range scalar fields
coupled to Dark Matter (DM) have been investigated as a possible
source of EEP violations~\cite{Carroll2009}, which again provides a
very appealing route towards independent confirmation of DM, making it
more tangible than only a hypothesis for otherwise unexplained
astronomical observations.

\section{EEP in the Context of Physics Today}
\label{sec:EEPtoday}

\subsection{Cosmology Context}
\label{sec:cosmo}

One of the most important discoveries of the past decade has been that
the present Universe is not only expanding, but it is also
accelerating~\cite{Perlmutter1999, Riess1998, Ade2013}. Such a
scenario is problematic within the standard cosmological model, based
on GR and the Standard Model of particle physics. Together, these two
models provide a set of predictions well in agreement with
observations: the formation of light elements in the early Universe
--- the big bang nucleosynthesis (BBN), the existence of the cosmic
microwave background (CMB), and the expansion of the
Universe. However, now the big challenge of modern cosmology and
particle physics is to understand the observed acceleration of the
Universe. Observations indicate that the content of matter and energy
in our Universe is about 68.3 \% dark energy (DE), 26.8 \% dark matter
(DM), and 4.9 \% baryonic matter~\cite{Ade2013}. These values are
obtained assuming the $\Lambda$-CDM model. There are independent
measurements of the DE component of the Universe from observations of
high red-shift Type Ia Supernovae~\cite{Perlmutter1999, Riess1998,
  Garnavich1998, Knop2003}. The evidence for DM comes essentially from
the analysis of galactic rotation curves~\cite{Faber1979}, acoustic
oscillations in the CMB~\cite{Hu1995, Jungman1996, Zaldarriaga1997},
large scale structure formation~\cite{Eisenstein1998, Eisenstein2005},
and gravitational lensing~\cite{Clowe2006, Zhang2007}. Nevertheless,
although there is such a strong evidence for the existence of DE and
DM, almost nothing is known about their nature and properties.

The simplest explanation of DE is the existence of a small, but
non-zero, cosmological constant $\Lambda$. The latter does not undergo
a dynamical evolution, and is conventionally associated to the energy
of the vacuum in a quantum field theory. In other words, the
cosmological constant is a constant energy density filling space
homogeneously and isotropically, and is equivalent physically to
vacuum energy. As a consequence, it should store the energy density of
the present day Universe and its value should be of the order of the
critical density. In fact, from the observations it follows that
$\Lambda \simeq H_0^2$, where $H_0 = 2\times 10^{-42}\,\text{GeV}$ is
the present value of the Hubble parameter and is related to the
dimension of the Universe. The vacuum energy density associated to the
cosmological constant is therefore $\rho_\Lambda = \Lambda/8\pi G
\simeq 10^{-47}\,\text{GeV}^4$ ($\simeq\rho_\text{critical}$). On the
other hand, arguments from quantum field theory imply that the vacuum
energy density is the sum of zero point energy of quantum fields with
a cutoff determined by the Planck scale ($m_\text{P} \simeq 1.22
\times 10^{19}\,\text{GeV}$) giving $\rho_\text{vacuum} \simeq
10^{74}\,\text{GeV}^4$, which is about 121 orders of magnitude larger
than the observed value. A lower scale, fixed for example at the QCD
scale, would give $\rho_\text{vacuum} \simeq 10^{-3}\,\text{GeV}^4$
which is still much too large with respect to $\rho_\Lambda$. From a
theoretical point of view, at the moment, there is no explanation as
to why the cosmological constant should assume the correct value at
the scale of the observed Universe. The only argument we can give is
based on the anthropic principle, \textit{i.e.} the idea that much
larger values would not have lead to the formation of stars, planets
and ultimately humans.

Rather than dealing directly with the cosmological constant to explain
the accelerating phase of the present Universe, a number of
alternative approaches and models have been proposed in the last
years. Some of these models are briefly summarized below:
\begin{enumerate}
\item Quintessence models~\cite{Wetterich1988, Ratra1988, Carroll1998}
  --- These models invoke a time evolving scalar field with an
  effective potential that provides the observed inflation;
\item Chameleon fields~\cite{Khoury2004a, Khoury2004b, Brax2004} ---
  In this model the scalar field couples to the baryon energy density
  and is homogeneous, varying across space from solar system to
  cosmological scales;
\item $K$-essence~\cite{Chiba2000, ArmendarizPicon2000,
  ArmendarizPicon2001} --- Here the scalar field sector does contain a
  non-canonical kinetic term;
\item Modified gravity arising out of string theory~\cite{Dvali2000}
  --- In this model the feedback of non-linearities into the evolution
  equations can significantly change the background evolution leading
  to acceleration at late times without introducing DE;
\item Chaplygin gases~\cite{Kamenshchik2001, Bilic2002, Bento2002} ---
  This model attempts to unify DE and DM in a unique setting, by
  allowing for a fluid with an equation of state which evolves between
  the two;
\item $f(R)$-gravity~\cite{Capozziello2011, Nojiri2007} --- In this
  model one considers instead of the Einstein-Hilbert action a generic
  function of the scalar curvature $R$, not necessarily linear in $R$
  as in the conventional GR. $f(R)$-gravity contains many features
  which make these models very attractive, as for example: (i) they
  provide a natural unification of the early-time inflation and the
  later-time acceleration of the Universe owing to the different role
  of the gravitational terms relevant at small and large scales; (ii)
  they allow to unify DM and DE; (iii) they provide a framework for
  the explanation of the hierarchy problem and unification of Grand
  Unified Theories (GUT) with gravity. However, some $f(R)$-models of
  gravity are strongly constrained (or ruled out) by solar system
  tests restricting the possible models;
\item Phantom Dark Energy~\cite{Caldwell2002}.
\end{enumerate}
Many of the models proposed in the literature are characterized by the
fact that a scalar field (or more than one scalar field) coupled or
not to gravity and ordinary matter is included in the action of
gravity.

On a fundamental ground, there are several reasons to introduce a
scalar field in the action describing gravity. A scalar field coupled
to gravity is an unavoidable aspect of all theories aimed at unifying
gravity with the other fundamental forces. These theories include
Superstring, Supergravity (SUGRA), $\text{M}$-theories. Moreover,
scalar fields appear both in particle physics and cosmology: the Higgs
boson in the Standard Model, the (string) dilaton entering the
supermultiplet of the higher dimensional graviton, the super-partner
of spin-$\frac{1}{2}$ in SUGRA. It also plays a non-trivial role in models
based on composite boson condensates. The introduction of a scalar
field gives rise typically to a violation of the EEP depending on its
coupling to the Lagrangian describing ordinary matter.

The above considerations apply to most extended theories of gravity
(scalar-tensor theories, $f(R)$-gravity, \textit{etc.}), which leaves
the foundation of relativistic gravity on a rather shaky ground. That
becomes a problem especially when trying to isolate the fundamental
properties of classical gravity which should be preserved in
approaches to quantum or emergent gravity~\cite{Capozziello2011}. The
STE-QUEST experiment will therefore play a crucial role not only for
searching possible violation of the EEP, but will shed light also on
what effective theories of gravity among those above mentioned is the
true theory for describing gravity.

A violation of the EEP in the \textit{dark sector} (DM and DE), comes
also from a possible coupling of DM to a scalar field. More precisely,
a (light) scalar field coupled to DM could mediate a long-range force
of strength comparable to gravity~\cite{Carroll2009, Carroll2010,
  Bean2008}. This kind of investigation is also motivated by the fact
that such interactions could account for features related to the DM
distribution as well as to DM-quintessence
interactions~\cite{Carroll2009, Carroll2010, Bean2008, Damour1990,
  Bertolami2005, Gubser2004}. Limits on such a force have been derived
from observations of DM dynamics in the tidal stream of the
Sagittarius dwarf galaxy which yields a force with strength less that
20 \% of gravity for a range of 20 $\text{kpc}$~\cite{Kesden2006a,
  Kesden2006b}. Moreover, as noted in Refs.~\cite{Carroll2009,
  Carroll2010, Bean2008}, if a new long-range force will be detected
in future, then it would be a signal of the presence of a new mass
hierarchy between the light scalar mass $m_\phi \lesssim
10^{-25}\,\text{eV}$ and the weak scale $m_\text{W} \simeq
10^2\,\text{GeV}$, in addition to the one between the weak scale and
the Planck scale. The possibility that a scalar field couples to
Standard Model particles implies that the force acting on ordinary
matter could be composition-dependent~\cite{Damour1996}. As a
consequence, such forces are tightly constrained by E\"otv\"os
experiments looking for violations of the weak
EEP~\cite{Schlamminger2008}. On the other hand, even if $\phi$ has
only an elementary (\textit{i.e.}, renormalizable) coupling to DM,
interactions between DM and ordinary matter will still induce a
coupling of $\phi$ to ordinary matter~\cite{Carroll2009, Carroll2010,
  Bean2008}. This can be thought of as arising from the scalar
coupling to virtual DM particles in ordinary atomic nuclei. Hence, a
fifth force coupled to the Standard Model is naturally expected in the
case in which a light scalar couples to a DM field having Standard
Model interactions.

Without any doubts, the equivalence of gravitational mass and inertial
mass represents one of the most fundamental postulates in
Nature. Theoretical attempts to connect GR to the Standard Model of
particles are affected by a violation of the
EEP~\cite{Damour1996}. Therefore, tests of the EEP turn out to be
important tests of unification scale physics far beyond the reach of
traditional particle physics experiments. The discoveries of DM and DE
have provided strong motivation to extend tests of the EEP to the
highest precision possible. In this respect, the STE-QUEST experiment
will play a significant role.

\subsection{Particle Physics Context}
\label{sec:part}

In the previous section, it already became clear that the difficulties
of GR in cosmology are closely related to those in particle
physics. In particular, in a quantum field theory (like the Standard
Model of particle physics), one would expect that the vacuum energy of
the fundamental fields should be observed in its gravitational
consequences, especially on the large scale of the Universe. However,
there is a huge discrepancy (121 orders of magnitude, or at least 44
orders of magnitude if one assumes the QCD scale, see
Sec.~\ref{sec:cosmo}) between the observed vacuum energy density of
the Universe (dark energy) and the one expected from the Standard
Model of particle physics. This has been considered a major problem in
modern physics, even before the discovery of dark energy when the
``observed'' value of the cosmological constant (or vacuum energy) was
compatible with zero~\cite{Weinberg1989}. And one might argue that
this problem has become even worse since the discovery of the
accelerated expansion of the Universe, and the associated small
\textit{but non-zero} value of $\Lambda$, as now one requires a
mechanism that does not completely ``block'' the gravitational effect
of vacuum energy, but suppresses it by a huge factor, \textit{i.e.}
some extreme fine tuning mechanism is required that is difficult to
imagine.

Another conceptual problem is that the Standard Model of particle
physics requires a number of dimensionless coupling constants to be
put in by hand, which seems somewhat arbitrary and is not very
satisfactory~\cite{Damour2012}. One of the aims of theoretical
developments is then to replace these constants by some dynamical
fields that provide the coupling constants (\textit{e.g.} moduli
fields in string theory, dilaton, \textit{etc.}), similarly to the
Higgs field giving rise to the mass of fundamental particles. As a
consequence the coupling constants become dynamical quantities that
vary in space-time (\textit{e.g.} space-time variation of the fine
structure constant $\alpha$), which necessarily leads to violations of
the EEP (violation of LPI, but also of WEP/UFF and LLI). However, the
resulting phenomenological consequences are such that in most
approaches one requires some mechanism to stabilize these fields in
order to be compatible with present-day constraints from EEP
tests~\cite{Taylor1988, Damour1994}. Although no firm predictions
exist, this makes the discovery of the effect of such fields
(\textit{e.g.} EEP violation) a distinct
possibility~\cite{Damour2012}.

The recent discovery of the Higgs particle at LHC confirms the
existence of the first fundamental scalar field, at least fundamental
down to the scale probed by the Standard Model. As discussed in the
previous section, scalar fields are ubiquitous in cosmology because
they easily provide a diffuse background: they play a central role in
most models of inflation or dark energy. It is thus important to have
identified at least one fundamental scalar field. There has been
attempts to make the Higgs field itself play a role in cosmology, by
coupling it to the curvature of space-time. This is for example the
model of Higgs inflation~\cite{Bezrukov2008}. At first glance, this
might seem to lead to violations of the equivalence principle but,
going to an Einstein frame, this gives rise to nonlinear interactions
of the Higgs field, which are down by powers of the Planck mass (or,
more precisely, $M_\text{P}/\xi$ if $\xi$ is the coupling of the Higgs
to curvature)~\cite{Bezrukov2011}.

Even if one disregards gravity, the Standard Model of particle physics
still does not address all the fundamental questions: in particular,
whereas it attributes the origin of mass to the Higgs non-vanishing
vacuum value, it does not explain the diversity of the masses of the
fundamental particles, \textit{i.e.} it does not explain the diversity
of the couplings of the matter to the Higgs field. One thus has to go
to theories beyond the Standard Model in order to answer these
questions. Most of these theories make heavy use of scalar fields, the
most notable examples being supersymmetry, which associates a scalar
field to any spin-$\frac{1}{2}$ matter field, string theory and
higher-dimensional theories. Some of these scalar fields may be
extremely light, or even massless, which leads to new types of long
range forces, and thus potential EEP violations, unless these fields
are universally coupled, a difficult property to achieve.

Moreover, the values of these scalar fields often have a predictive
role in setting the value of fundamental constants or ratios of mass
scales. Because they are weakly coupled to ordinary matter, they may
not have reached their fundamental state, in which case they are still
evolving with time. This leads to a time dependence of the
corresponding constants or mass scales, and thus again to a potential
violation of the equivalence principle.

\section{Quantum Mechanics and the EEP}
\label{sec:quantum}

Quantum tests of the Equivalence Principle differ from classical ones
because classical and quantum descriptions of motion are fundamentally
different. In particular, the Universality of Free Fall (or WEP) has a
clear significance in the classical context where it means that
space-time trajectories of test particles do not depend on the
composition of these particles. How UFF/WEP is to be understood in
Quantum Mechanics is a much more delicate point. The subtlety of
discussions of the EEP in a quantum context is also apparent in the
debate about the comparison of various facets of the EEP, in
particular the UFF and the LPI~\cite{Muller2010, Wolf2011,
  Giulini2012}. More generally, considering quantum phenomena in the
context of gravity poses many conceptual and fundamental difficulties
as discussed below. Although not all of these are directly explored by
STE-QUEST, they provide a broad picture of the limits of our knowledge
in this domain and thus the interest of experiments like STE-QUEST
that have the discovery potential for expected and unexpected results
that might shed light on this frontier of physics.

\begin{figure}[t]
\begin{center}
\begin{tabular}{c}
\includegraphics[width=15cm]{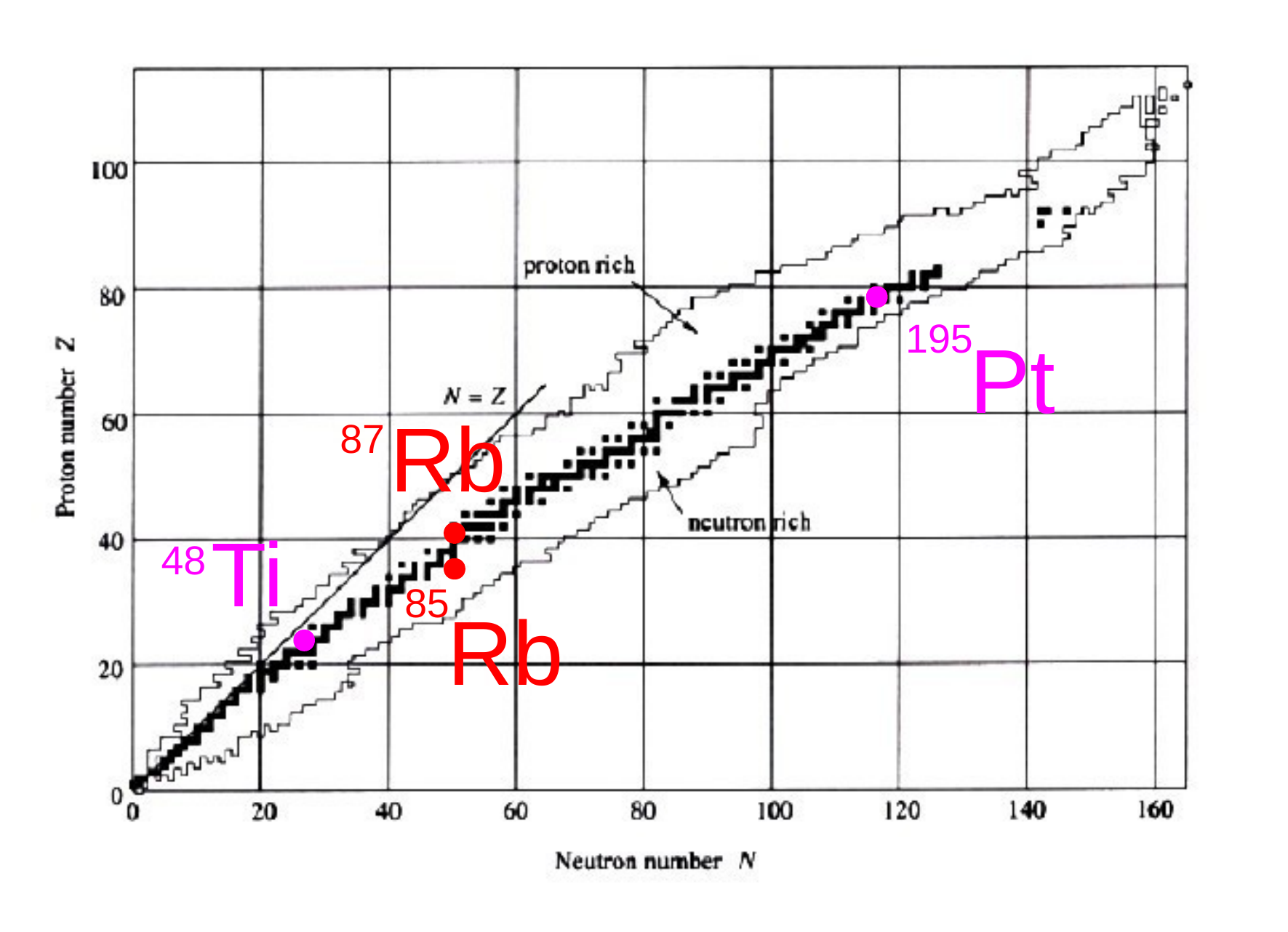}
	\end{tabular}
\caption{The valley of stable nuclei in the $(N,Z)$ plane, with the
  nuclei used in STE-QUEST in red, and those of $\mu$-SCOPE in violet.}
\label{fig:nuclei}
\end{center}
\end{figure}

Let us first discuss the case where no distinction is made between
classical and quantum tests, by evaluating different UFF tests with
respect to non-standard theories, as was done for example by Damour \&
Donoghue~\cite{Damour2010} for the specific case of couplings to a
light dilaton. The same type of argument is also valid for a vector
field like the $U$ boson~\cite{Fayet1986, Fayet1990}. In these cases a
similar value for the fractional inaccuracy (\textit{e.g.} $10^{-15}$)
leads to a larger sensitivity for the free fall of titanium
(${}^{48}\text{Ti}$) and platinum (${}^{196}\text{Pt}$) test masses
(\textit{e.g.} the $\mu$-SCOPE mission) than for STE-QUEST where two
isotopes of rubidium (${}^{85}\text{Rb}$ and ${}^{87}\text{Rb}$) are
compared. To be more quantitative, the sensitivity depends on the
difference in $E_i/M$ ratios between the two test masses, with $E_i$
being a particular type of nuclear binding energy. The difference in
sensitivities then simply stems from the locations of these atoms
along the sequence of stable heavy elements in the $(N,Z)$ plane as
shown in Fig.~\ref{fig:nuclei}.

Let us be more quantitative on the comparison STE-QUEST versus
$\mu$-SCOPE based on Fig.~\ref{fig:nuclei}. In a model such as in
Ref.~\cite{Damour2010}, the nuclear binding energy of nuclei is the
sum of several contributions ($A=N+Z$):
\begin{equation}\label{Energy}
E = - 16.0\,A + 17.0\,A^{2/3} + 23.0\,\frac{(N-Z)^2}{A} +
0.7\,\frac{Z(Z-1)}{A^{1/3}} -
6.0\,\frac{(-1)^N+(-1)^Z}{A^{1/2}}~\text{MeV}\,.
\end{equation}
These are, respectively, the volume energy $\propto A$, the surface
energy $\propto A^{2/3}$, the asymmetry energy, the (repulsive)
Coulomb energy and the pairing energy. In a model for the coupling of
the dilaton to fundamental matter fields (quarks and gluons), one
expects that all the terms in the formula~\eqref{Energy} will
contribute separately to the violation of the EEP. For two atoms A and
B and a given type of energy $i$ in Eq.~\eqref{Energy} we pose
\begin{equation}\label{deltai}
\delta_i(\text{A},\text{B}) =
\frac{E_i(\text{A})}{M(\text{A})}-\frac{E_i(\text{B})}{M(\text{B})}\,.
\end{equation}
Then an indicator of the uncertainty with which STE-QUEST and
$\mu$-SCOPE (both having fractional inaccuracy of the order of
$10^{-15}$) can test each of the separate terms in Eq.~\eqref{Energy}
is provided by the ratio
\begin{equation}\label{si}
s_i(\text{A},\text{B}) =
\frac{10^{-15}}{\delta_i(\text{A},\text{B})}\,.
\end{equation}
\begin{table*}[t]
\begin{center}
\begin{tabular}{|r|r|r|r|r|} 
\hline & $E_i({}^{195}\text{Pt})/\text{MeV}$ &
$E_i({}^{48}\text{Ti})/\text{MeV}$ &
$\delta_i({}^{195}\text{Pt},{}^{48}\text{Ti})$ &
$s_i({}^{195}\text{Pt},{}^{48}\text{Ti})$\\ \hline \hline
\textbf{volume} & $-3120.0$ & $-768.0$ & $1.4\times 10^{-6}$ &
$7.3\times 10^{-10}$ \\ \hline \textbf{surface} & $571.7$ & $224.5$ &
$-1.9\times 10^{-3}$& $-5.4\times 10^{-13}$ \\ \hline
\textbf{asymmetry} & $179.4$ & $7.7$ & $8.1\times 10^{-4}$& $1.2\times
10^{-12}$ \\ \hline \textbf{Coulomb} & $725.0$ & $89.0$ & $2.0\times
10^{-3}$& $5.0\times 10^{-13}$ \\ \hline \textbf{pairing} & $0.0$ &
$-1.7$ & $3.8\times 10^{-5}$& $2.6\times 10^{-11}$ \\ \hline\hline
\textbf{total} & $-1643.9$ & $-448.5$ & $9.7\times 10^{-4}$&
$1.0\times 10^{-12}$ \\ \hline
\end{tabular}
\caption{Sensitivity of the pair
  ${}^{48}\text{Ti}$-${}^{196}\text{Pt}$ to various nuclear
  energies.}\label{tab:pairTiPt}
\end{center}
\begin{center}
\begin{tabular}{|r|r|r|r|r|} 
  \hline & $E_i({}^{87}\text{Rb})/\text{MeV}$ &
  $E_i({}^{85}\text{Rb})/\text{MeV}$ &
  $\delta_i({}^{87}\text{Rb},{}^{85}\text{Rb})$ &
  $s_i({}^{87}\text{Rb},{}^{85}\text{Rb})$\\ \hline \hline
  \textbf{volume} & $-1392.0$ & $-1360.0$ & $2.3\times 10^{-7}$ &
  $4.3\times 10^{-9}~$ \\ \hline \textbf{surface} & $333.8$ & $328.6$
  & $-3.2\times 10^{-5}$& $-3.1\times 10^{-11}$ \\ \hline
  \textbf{asymmetry} & $44.7$ & $32.7$ & $1.4\times 10^{-4}$&
  $7.3\times 10^{-12}$ \\ \hline \textbf{Coulomb} & $210.4$ & $212.1$
  & $-8.1\times 10^{-5}$& $-1.2\times 10^{-11}$ \\ \hline
  \textbf{pairing} & $0.0$ & $0.0$ & $0.0$& $\infty$ \\ \hline\hline
  \textbf{total} & $-803.1$ & $-786.5$ & $2.4\times 10^{-5}$&
  $4.2\times 10^{-11}$ \\ \hline
\end{tabular}
\caption{Sensitivity of the pair ${}^{85}\text{Rb}$-${}^{87}\text{Rb}$
  to various nuclear energies.}\label{tab:pairRu}
\end{center}
\end{table*}
The results are tabulated in Tables~\ref{tab:pairTiPt}
and~\ref{tab:pairRu}. As we see the pair
${}^{48}\text{Ti}$--${}^{196}\text{Pt}$ is between a factor $\sim 5$
and a factor $\sim 60$ times more sensitive (depending on which $E_i$
is considered) than the pair
${}^{85}\text{Rb}$--${}^{87}\text{Rb}$. In addition the pair
${}^{85}\text{Rb}$--${}^{87}\text{Rb}$ has no sensitivity to the
pairing energy because the difference of the number of neutrons in the
two isotopes is even.

However, from a wider phenomenological point of view, notice that the
${}^{85}\text{Rb}$--${}^{87}\text{Rb}$ test has to be considered as
exploring a variation in the table of nuclei which is orthogonal and
complementary to that along the main sequence, mainly tested by the
pair ${}^{48}\text{Ti}$--${}^{196}\text{Pt}$. In this context, the
precision of STE-QUEST has to be compared to existing tests between
two $\text{Rb}$ isotopes rather than to $\mu$-SCOPE. This is similar
to methods used in the context of particle physics where the CERN
Scientific Council has accepted experiments~\cite{Doser2010,
  Perez2012} for testing the free fall of cold anti-hydrogen atoms at
a level of the order of $10^{-2}$. With most non-standard theories
used to compare the interest of experiments, the targeted precision is
far from what is already known from classical tests of the EEP. The
fact that the experiments have been judged to be worthy shows the
peculiar interest of tests of UFF performed with non-classical objects
like antimatter or quantum objects.

Let us now discuss a number of physical hypotheses one is implicitly
making when doing the above classical comparison, simply looking at
the different locations of the tested materials in
Fig.~\ref{fig:nuclei}, \textit{i.e.} assuming that there is nothing
special about quantum tests.

The first implicit assumption is that Quantum Mechanics is valid in
the freely falling frame associated with classical test bodies in the
definition of WEP. Indeed, the usual definition of the EEP states that
special relativity holds in the freely falling frame of WEP without
reference to quantum mechanics.\footnote{Recall that relativistic
  quantum mechanics did not exist at the time of the earliest
  formulation of the equivalence principle by Einstein.} Of course,
this extension of the notion of freely falling frame to quantum
mechanics is always implicit and ``obvious''. It is used when one
computes the phase shift of a matter wave interferometer in a gravity
field, using the full machinery of quantum mechanics, for instance
Feynman's path integral formalism~\cite{Storey1994}.

Another important implicit assumption is that any possible violation
of the EEP must be due to a new fundamental interaction (which
superposes to the gravitational force), and that fundamental
interactions are described in the framework of quantum field theory
(QFT) by bosonic fields. In particular, the formalism of QFT must be
true for that field, \textit{e.g.} the procedures of second
quantization and of renormalization. The consequence is that the
violation of the EEP is either due to a scalar spin-0 field (dilaton)
or a vector field (for instance the $U$ boson~\cite{Fayet1986,
  Fayet1990}). Indeed, recall that higher-order spin fields
($s\geqslant 2$) yield difficulties, for instance the coupling of an
additional spin-2 field to the metric spin-2 field of GR is
problematic. Of course there are theorems that fundamental
interactions in the framework of relativistic quantum mechanics
necessarily involve the notion of fields, but as physicists we also
want to prove our theorems experimentally.

The previous statements represent the state-of-the-art of Physics that
we have today; if one of these would turn out to be wrong this would
provoke a major crisis in Physics. Nevertheless, because they are so
fundamental, these statements are worth being experimentally verified
wherever possible. It is true that they are tested every day in
particle accelerators, but in a regime where the gravitational field
plays essentially no role. Testing the EEP for quantum waves in the
presence of gravity represents a new way of testing some of our
deepest beliefs in Physics at the interplay between QM and GR.

Additionally, there are a number of other concerns regarding the
quantum to classical comparison in general, which illustrate the
difficulties in this region of Physics and thus the interest of any
experimental guidance.

The variety of quantum states is much larger than that of classical
ones and it seems therefore plausible that quantum tests may
ultimately be able to see deeper details of couplings between matter
and gravity than classical ones. When considering non-standard
couplings of matter to gravity, there might be a difference between
how the wave packet centre is moving and how it is
deforming~\cite{Kajari2010, Bourdel2011, Unnikrishnan2012}. As an
illustration in a concrete example, let us consider the free fall in a
gravitational field of a particle in QM described by the wave function
$\Psi$. We assume that the wave function is initially
Gaussian. Schr\"odinger's equation with Hamilton operator
\begin{equation}\label{Hamilton}
\hat{H}=\frac{\hat{p}_z^2}{2m}+m g \hat{z}\,
\end{equation}
is satisfied, where the second term is the usual Newtonian
gravitational potential. We compute the time of flight of this
particle from some initial position $z_0$ up to $z=0$, the initial
position being determined by the expectation value $z_0 =
\langle\hat{z}\rangle_{\Psi_0}$ of the position in the Gaussian
initial state $\Psi_0$. The time of flight is statistically
distributed with the mean value agreeing with the classical universal
value,
\begin{equation}\label{T}
T = \sqrt{\frac{2 z_0}{g}}\,.
\end{equation}
However, the standard deviation of the measured values of the time of
flight around $T$ depends on the mass of the particle
\begin{equation}\label{sigma}
\sigma = \frac{\hbar}{\Delta_0 \,m g}\,,
\end{equation}
where $\Delta_0$ is the width of the initial Gaussian wave packet. In
this sense the quantum motion of the particle is non-universal, as it
depends on the value of its mass~\cite{Lammerzahl1996, Viola1997,
  Davies2004}.

Another example is the role of intrinsic spin of quantum probes, that
has no classical equivalent. For classical particles, the EEP is
implemented by the rule of the minimal coupling (see \textit{e.g.}
Ref.~\cite{Weinberg1972}): in the presence of the gravitational field
we replace the Minkowski metric $\eta_{\alpha\beta}$ in the Lagrangian
of special relativity by the curved space-time metric $g_{\mu\nu}$ of
GR. Suppose that a classical body is made of $N$ particles with
positions $\mathbf{x}_a$ and velocities $\mathbf{v}_a$ interacting
through the classical electromagnetic field $A_\alpha$, with dynamics
resulting from the Lagrangian
\begin{equation}
L_\text{SR} = L[\mathbf{x}_a, \mathbf{v}_a, A_\alpha,
  \eta_{\alpha\beta}]\,
\end{equation}
in special relativity. Then the Lagrangian describing the dynamics of
this body in GR will simply be
\begin{equation}
L_\text{GR} = L[\mathbf{x}_a, \mathbf{v}_a, A_\mu, g_{\mu\nu}]\,.
\end{equation}
The procedure to couple a quantum field to gravity is much more
complex and, we argue, more fundamental than for the coupling of
classical fields. The Lagrangian of the quantum field (\textit{e.g.}
the Dirac field) depends on the derivative of the field because of the
intrinsic spin, and requires additional formalisms like tetrads and
the spinorial representations of the Lorentz group and the associated
spinorial derivative. So, while classical matter is coupled to gravity
by using only the metric ($\eta_{\alpha\beta}$ replaced by
$g_{\mu\nu}$), quantum fields associated with electrons and other
fermions are coupled to gravity through tetrads, which may be
considered as a deeper representation of space-time (the metric is
immediately deduced from the tetrad, but the inverse is not true),
with a more complicated formalism.

Of course, atom interferometry tests of the EEP are usually performed
with unpolarized spin states ($m_F = 0$) because the latter are
insensitive to magnetic fields at first order. They can also be
performed with other Zeeman sublevels (spin polarized states with $m_F
\neq 0$) with a somewhat reduced precision due to the first-order
coupling with magnetic fields. This possibility of performing
spin-dependent tests is an obvious advantage of quantum tests over
classical versions of EEP tests, though the latter ones may of course
perform tests with spin-polarized matter~\cite{Hoedl2011}. Comparison
of these various tests could be done by following the line already
opened for spin-dependent clock measurements~\cite{Wolf2006}, using
the Standard Model Extension (SME) framework~\cite{Kostelecky2011a}.

Our main concern about the frontier between QM and GR is of course the
absence of a consistent quantum theory of gravity.  Its
non-renormalisability within standard perturbative methods has led to
a variety of suggestions and alternative approaches, the most
pragmatic being to incorporate the gravitational field into the
effective-field-theory (EFT) program, which results in definite
prescriptions for the computations of low-energy quantum corrections
at scales well above the Planck scale~\cite{Burgess2004}. In
particular, it allows for computations of metric fluctuations in
inflationary cosmology and subsequent applications to explain CMB
anisotropies. On the other hand, taking the geometric interpretation
of gravity as the central paradigm, it has been suggested that gravity
will eventually defy standard quantization approaches and that the
sought-for reconciliation will impose at least as much change on our
present notion of Quantum(-Field) Theory as it does on classical
GR~\cite{Penrose2014}. In view of these diverging attitudes it is
important to note that they will already differ in their respective
answers to the most mundane physical questions, such as: what is the
gravitational field sourced by a quantum system in a state represented
by a wave function $\Psi$?  Here the so-called semi-classical theory
comes into play, which states that the gravitational field obeys the
Einstein field equation
\begin{equation}\label{semiclass}
  G^{\mu\nu} = \frac{8\pi G}{c^4}\,\langle
  \hat{T}^{\mu\nu}\rangle_\Psi\,,
\end{equation}
where the expectation value of the stress-energy quantum operator
$\hat{T}^{\mu\nu}$ in the given quantum state $\Psi$ replaces the
classical stress-energy tensor $T^{\mu\nu}$.

The obvious and orthodox interpretation of this equation is that of an
approximation to quantum gravity for states $\Psi$ producing
negligible fluctuations in the sourcing stress-energy.  If we assume
quantum gravity to obey the usual rules of measurement and
quantum-state reduction, then it is easy to see that~\eqref{semiclass}
cannot be a valid description of such
processes~\cite{Wald1984}. Indeed, suppose that we have a state of
matter made of the superposition of two states (each with probability
$1/2$) in which the matter is localized, respectively, into two
different disjoint regions. Then, according to Eq.~\eqref{semiclass},
the gravitational field will be generated by half the matter in the
first region and by half the matter in the other distinct region. This
is clearly incompatible with a measurement of the location of the
matter (and the associated collapse of the wave function), since after
measurement all the matter will be either located entirely in the
first, or entirely in the second region.

On the other hand, if we allow for the possibility that quantum
gravity will imply changes to the usual rules of quantum-state
reduction, \textit{e.g.}, along the lines of Ref.~\cite{Penrose2014},
then Eq.~\eqref{semiclass} might acquire a more fundamental status. In
fact, the first suggestion of Eq.~\eqref{semiclass} was made in
connection with the logical possibility that gravity is not to be
quantized at all~\cite{Rosenfeld1963}. Based on this, it has further
been suggested to experimentally scrutinize the alleged necessity of
quantum gravity, \textit{e.g.}, by looking at situations in which the
classical gravitational self-field computed according
to~\eqref{semiclass} affects the quantum-mechanical dispersion as a
result of the non-linearities that the dependence of the spacetime
metric on the state $\Psi$ [resulting from~\eqref{semiclass}]
introduces into the quantum dynamical equations for $\Psi$ (which in
turn depend on the metric)~\cite{Carlip2008, Giulini2011}.

Back to the orthodox viewpoint, where Eq.~\eqref{semiclass} is only of
approximate validity, we maintain that it can be an effective
description of quantum systems under gravity, including their own
field, and as such it makes sense and has interesting physical
consequences, like the back-reaction of Hawking's radiation on black
holes, or that of quantum fields in the early Universe. Although the
semi-classical theory will not be checked directly by quantum tests of
the EEP, the above issues and paradoxes remind us that we do not
dispose of a consistent quantum theory of gravity and that
experimental evidence exploring the relationship between QM and GR is
direly needed.

In summary, although there is no established theory that favors
quantum tests of the EEP, there are nonetheless a number of
difficulties in the frontier between QM and GR due to the absence of a
quantum theory of gravitation, that call for experiments lying at that
frontier, like quantum tests of the EEP proposed by STE-QUEST.

On the experimental side, quantum mechanical tests have to be
considered as opening a new technological avenue based on quantum
sensors, which is probably the best solution for future much improved
tests. Whereas macroscopic tests approach their ultimate limits after
years of scientific research and technical development, this is not
the case for atomic tests. In particular, the accuracy of the atom
interferometry test in STE-QUEST ($2 \times 10^{-15}$) promises an
improvement by 8 orders of magnitude over the best existing test
($10^{-7}$) between two Rb isotopes. With this improvement, it already
would reach the level of accuracy of $\mu$-SCOPE~\cite{Touboul2001},
while still having possibilities for future improvements.

\section{STE-QUEST Tests of the Einstein Equivalence Principle}
\label{sec:STEtest}

In this section, we discuss specifically the tests of the EEP carried
out by STE-QUEST. We first describe a general theoretical framework
for the WEP/UFF and LPI tests that allows a classification and
comparison of the different experiments and clarifies the
complementarity between the different types of tests. We then use that
framework to compare each of the planned STE-QUEST experiments to
existing and expected measurements in the same domain and point out
the improvements expected from STE-QUEST. Finally we address the
possible STE-QUEST tests of Lorentz Invariance and CPT symmetry using
another theoretical framework particularly adapted for that purpose.

\subsection{Different Tests of the EEP}
\label{sec:diffEEP}

Tests of the different aspects of EEP (\textit{i.e.} WEP, LLI, and
LPI), and the relations between them, are best discussed within the
``modified Lagrangian framework'', which is a powerful formalism
allowing deviations from GR and metric theories of gravity, but at the
same time permitting a coherent analysis of various
experiments~\cite{Nordtvedt1975, Haugan1979, Will1993, Wolf2011}. The
formalism describes a large class of non-metric theories in a way
consistent with Schiff's conjecture and energy conservation. This
class of theories is defined by a single Lagrangian, in which the
coupling between gravitation and different types of mass-energies is
generically not universal. In a simplified variant of the formalism,
we consider a composite body of mass $m$ (\textit{e.g.} an atom in a
STE-QUEST experiment) in the Newtonian gravitational potential
$U(\mathbf{x}) = GM/r$ of the Earth, where $r = \vert\mathbf{x}\vert$,
thus obeying the Lagrangian
\begin{equation}\label{L}
L = -m\,c^2+m\,U +\frac{1}{2}m \,\mathbf{v}^2\,.
\end{equation}
We postulate that the mass $m=m(\mathbf{x})$ of this body depends on
the position $\mathbf{x}$ through a violation of the LPI. This is
modelled by assuming that a particular internal energy of the body,
$E_X=E_X(\mathbf{x})$, behaves anomalously in the presence of the
gravitational field, where $X$ refers to the type of interaction
involved (electromagnetic, nuclear, spin-spin, spin-orbit,
\textit{etc.}). For simplicity, because we have in mind the discussion
of the red-shift test versus the test of the UFF, both being performed
by STE-QUEST, we consider only a dependence on $\mathbf{x}$ to model
the violation of the LPI. It could be possible to include also a
dependence on the velocity $\mathbf{v}$ to model a violation of
LLI. Separating out $E_X(\mathbf{x})$ from the other forms of energies
$\overline{E}_Y$ composing the body and which are supposed to behave
normally, we write
\begin{equation}\label{mx}
m(\mathbf{x}) = \overline{m} + \frac{1}{c^2}\biggl[E_X(\mathbf{x}) +
  \sum_{Y\not=X} \overline{E}_Y\biggr]\,.
\end{equation}
Here $\overline{m}$ denotes the sum of the rest masses of the
particles constituting the body; $\overline{m}$ and all
$\overline{E}_Y$'s are constant. The violation of LPI is modeled in
the simplest way by assuming that at the leading order
\begin{equation}\label{EX}
E_X(\mathbf{x}) = \overline{E}_X + \beta^{(a)}_X \,\overline{m}
\,\Delta U(\mathbf{x})\,,
\end{equation}
where $\Delta U=U_\oplus-U$ is the potential difference with respect
to some reference point, \textit{e.g.} the surface of the Earth. The
parameter $\beta^{(a)}_X$ is dimensionless and characterizes the
violation of LPI. It depends on the particular type of mass-energy or
interaction under consideration, \textit{e.g.} $\beta^{(a)}_X$ would
be different for the electromagnetic or the nuclear interactions, with
possible variations as a function of spin or the other internal
properties of the body, here labelled by the superscript $(a)$. Thus
$\beta^{(a)}_X$ would depend not only on the type of internal energy
$X$ but also on the type of body $(a)$. Defining now the ``normal''
contribution to the total mass,
\begin{equation}
m_0 = \overline{m} + \sum_Y \frac{\overline{E}_Y}{c^2}\,,
\label{mbar}
\end{equation}
and replacing $m(\mathbf{x})$ by its explicit expression into the
Lagrangian \eqref{L} we obtain
\begin{equation}\label{L1}
L = - m_0 \,c^2+ m_0\left(U - \beta^{(a)}_X \Delta U\right)
+\frac{1}{2}m_0 \mathbf{v}^2\,,
\end{equation}
where we have neglected higher-order terms, which are of no relevance
for the discussion.

We can now analyze the traditional free-fall and red-shift
experiments. By varying Eq.~\eqref{L1}, we obtain the equation of
motion of the body as
\begin{equation}\label{acc}
\frac{\ud\mathbf{v}}{\ud t} = \bigl(1+\beta^{(a)}_X\bigr)\bm{\nabla}
U\,,
\end{equation}
which shows that the trajectory is affected by the violation of LPI
and is not universal. In fact, we see that $\beta^{(a)}_X$ measures
the non-universality of the ratio between the body's passive
gravitational mass and inertial mass. Thus, in this framework, the
violation of LPI implies a violation of UFF (and WEP), and
$\beta^{(a)}_X$ is the WEP-violating parameter. This is a classic
proof of the validity of Schiff's conjecture~\cite{Schiff1960}.

The violation of LPI is best reflected in classical red-shift
experiments, which can be analysed using a cyclic Gedanken experiment
based on energy conservation. This was done in
Ref.~\cite{Nordtvedt1975}, extending a famous argument by Einstein
himself. The result for the fractional frequency shift $z$ in a Pound
\& Rebka-type experiment~\cite{Pound1960} is then
\begin{equation}\label{z}
z = \bigl(1+\alpha^{(a)}_X\bigr)\,\frac{\Delta U}{c^2}\,,
\end{equation}
where the LPI-violating parameter $\alpha^{(a)}_X$ is again
non-universal. The important point is that, within the framework of
the modified Lagrangian~\eqref{L1}, the LPI-violating parameter
$\alpha^{(a)}_X$ is related in a precise way to the WEP-violating
parameter $\beta^{(a)}_X$ (see Ref.~\cite{Nordtvedt1975}):
\begin{equation}\label{alphabeta}
\beta^{(a)}_X = \alpha^{(a)}_X\,\frac{\overline{E}_X}{\overline{m}
  \,c^2}\,.
\end{equation}
Therefore tests of LPI and WEP are not independent, and we can compare
their different qualitative meanings. Since for typical energies
involved we shall have $\overline{E}_X\ll\overline{m} \,c^2$, this
means that $\beta_X\ll\alpha_X$, where $\beta_X$ and $\alpha_X$ denote
some typical values of the parameters. For a given set of LPI and WEP
tests, their relative merit is given by Eq.~\eqref{alphabeta} and it
is dependent on the model used, \textit{i.e.} the type of anomalous energy
$E_X$ and the employed materials or bodies.

For example, let us assume a model in which all types of
electromagnetic energy are coupled in a non-universal way,
\textit{i.e.} $\beta_\text{EM} \neq 0$ (with all other forms of
energies behaving normally), and where the clock transition is purely
electromagnetic. The UFF test between two materials $(a)$ and $(b)$,
both containing electromagnetic energy (\textit{e.g.} binding energy),
is carried out with an uncertainty of $\vert\beta^{(a)}_\text{EM} -
\beta^{(b)}_\text{EM}\vert\simeq\vert\beta_\text{EM}\vert \lesssim
10^{-13}$ in best current experiments~\cite{Schlamminger2008}. On the
other hand the LPI test for a clock of type $(c)$ based on an
electromagnetic transition,\footnote{We assume in this example that
  electromagnetism plays the same role in the nuclear binding energy
  and the hydrogen hyperfine transition.} is carried out with an
uncertainty of $\vert\alpha^{(c)}_\text{EM}\vert
\simeq\vert\alpha_\text{EM}\vert \lesssim 10^{-4}$ in the GP-A
experiment~\cite{Vessot1979}. For macroscopic test bodies, the nuclear
electromagnetic binding energy contributes typically
$\overline{E}_\text{EM}/(\overline{m}c^2) \simeq 10^{-3}$ of the total
mass, so from Eq.~\eqref{alphabeta} we have $\vert\beta_\text{EM}\vert
\simeq 10^{-3}\,\vert\alpha_\text{EM}\vert$, which means that the WEP
test yields $\vert\alpha_\text{EM}\vert\lesssim 10^{-10}$, a much more
stringent limit than the red-shift test ($\vert\alpha_\text{EM}\vert
\lesssim 10^{-4}$).

However, that result depends on the particular model used. If we
assume another model in which the nuclear spin plays a role leading to
a non-universal coupling of atomic hyperfine energies, \textit{i.e.}
$\beta_\text{HF} \neq 0$ (with other forms of energies and properties
of the body behaving normally), the result is different. Atomic
hyperfine energies are of order $10^{-24}\,\text{J}$ (corresponding to
GHz transition frequencies), which for typical atomic masses leads to
$\bar{E}_\text{HF}/(\bar{m}c^2)\simeq 10^{-16}$. As a consequence, WEP
tests set a limit of only $\vert\alpha_\text{HF}\vert\lesssim 10^{3}$,
while LPI tests using hyperfine transitions (\textit{e.g.} H-masers) set a
limit of about $\vert\alpha_\text{HF}\vert \lesssim 10^{-4}$. The
conclusion is therefore radically different in this model where LPI
tests perform orders of magnitude better than WEP tests.

To summarize, the two types of tests, WEP (or UFF) and LPI
(red-shift), are complementary, and need to be all pursued, because
depending on the model used either one of the tests can prove
significantly more sensitive than the other. The main goal of
STE-QUEST is to perform at once the different types of tests of the
EEP, with good and in some case unprecedented precision: the WEP/UFF
test, which will be done by mean of atom interferometry, the
red-shift/LPI test through clock comparisons with optical and
microwave links, and also a test of LLI, whose comparison with the WEP
and LPI tests could be discussed in a way similar to what was
presented above.

\subsection{STE-QUEST Test of the Weak Equivalence Principle}
\label{sec:STEweak}

The atom interferometer (ATI) of STE-QUEST is described in detail in
the STE-QUEST Yellow Book~\cite{YellowBook} and in the
article~\cite{Aguilera2013}. Here we only recall the main principle
and some key numbers in the operation and measurements. The STE-QUEST
ATI is a dual-species atom interferometer using the two isotopes of
\text{Rb} (namely ${}^{85}\text{Rb}$ and ${}^{87}\text{Rb}$) which are
simultaneously trapped and cooled by a sequence involving atoms
manipulation by lasers and magnetic fields. Atoms are cooled to
temperatures below the critical temperature (a few $\text{nK}$) for
Bose-Einstein Condensation (BEC), which allows operation of the
interferometer with degenerate quantum gases (BECs). The complete
trapping and cooling process lasts about $10\,\text{s}$. The two
isotopes are then released into free fall and subject simultaneously
to a Mach-Zender interferometer. Each atom undergoes three laser
pulses that coherently split, reverse, and recombine the wave packets
during a time interval of $10\,\text{s}$. The actual separation of the
coherent wave packet parts during the interferometer sequence is up to
$10\,\text{cm}$ and larger than their respective size by more than two
orders of magnitude. During each of the pulses, the laser phase is
``imprinted'' onto the matter-wave phase, so that on recombination the
interference of the two waves (read out \textit{via} the populations
of internal states) provides the information on the acceleration of
the freely falling matter waves with respect to the laser source. The
final observable is then the difference between the measured
accelerations of the two isotopes, \textit{i.e.} the differential
acceleration of the ${}^{85}\text{Rb}$ and ${}^{87}\text{Rb}$ matter
waves.

The STE-QUEST ATI thus provides a test of the universality of free
fall or weak equivalence principle (UFF/WEP). Such tests are generally
quantified by the E\"otv\"os ratio $\eta_{AB}$ for two test objects
$A$ and $B$ and a specified source mass of the gravitational field:
\begin{equation}\label{etaAB}
\eta_{AB} = 2\,\frac{a_A - a_B}{a_A + a_B} = \beta_A - \beta_B \,,
\end{equation}
where $a_i$ ($i=A,B$) is the acceleration of object $i$ with respect
to the source mass and $\beta_i$ is the parameter introduced in
Eq.~\eqref{acc}. Note that for a given experiment the data can be
interpreted with respect to different source masses (see \textit{e.g.}
Ref.~\cite{Schlamminger2008}) with corresponding different results for
$\eta_{AB}$, and Eq.~\eqref{alphabeta} can be further refined in a
model-dependent way when searching violations linked to particular
types of mass-energy (see Sec.~\ref{sec:diffEEP}).
\begin{figure}[t]
\begin{center}
\begin{tabular}{c}
\includegraphics[width=16cm]{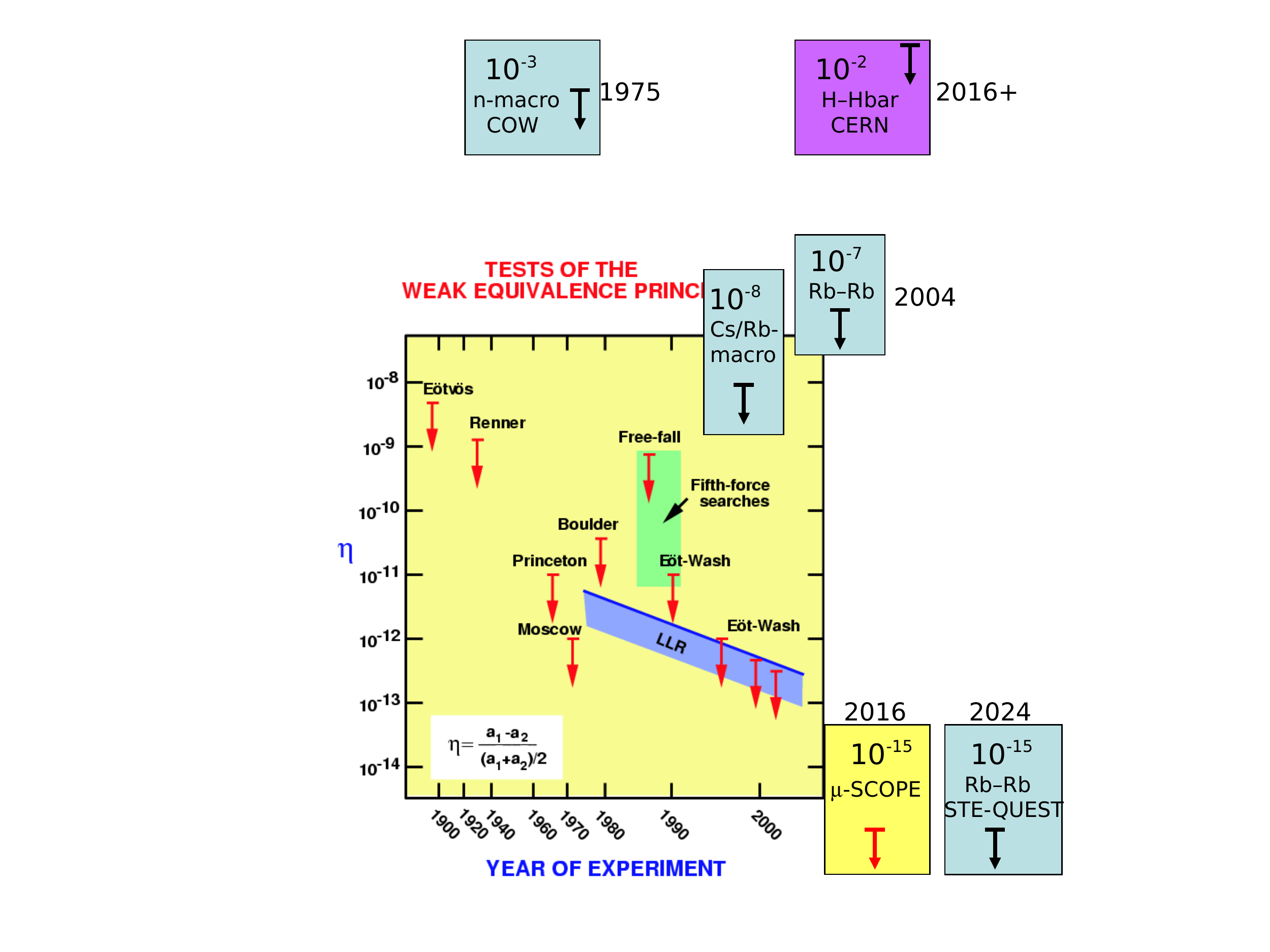}
	\end{tabular}
        \caption{Present and upcoming tests of WEP/UFF in the Earth
          field (except for LLR, which is in the Sun field) adapted
          from Ref.~\cite{Will2006}. Experiments using macroscopic
          test masses are represented by red arrows on a yellow
          background; experiments involving at least one quantum
          object by black arrows on a blue background. COW stands for
          the Collela-Overhauser-Werner experiment~\cite{Colella1975}
          using neutron interferometry compared to macroscopic test
          masses. $\text{Cs}/\text{Rb}$-macro~\cite{Peters1999,
            Merlet2010} are the similar atom interferometry
          experiments. For completeness the violet entry is the
          hydrogen \textit{vs.} anti-hydrogen test under construction
          at CERN.}\label{fig:testWEP}
\end{center}
\end{figure}

The useful ATI measurements are performed during about $0.5\,\text{h}$
around perigee, when the sensitivity is largest. When taking into
account all perturbing effects (gravity gradients, vibrations,
magnetic fields, \textit{etc.}) the single-shot ($20\,\text{s}$ cycle
time) sensitivity is about $2.9 \times 10^{-12}\,\text{m}/\text{s}^2$
in differential acceleration of the two isotopes, and is dominated by
the atomic shot noise. As a consequence, with the STE-QUEST baseline
orbit, the goal of $2 \times 10^{-15}$ sensitivity (statistical
uncertainty) in the E\"otv\"os ratio can be reached in less than 1.5
years (see~\cite{YellowBook, Aguilera2013} for details) with good
prospects for reaching $1 \times 10^{-15}$ in the mission
lifetime. Systematic effects are estimated to be below the $2 \times
10^{-15}$ level once calibrated, with the possibility of carrying out
some of the calibrations during the rest of the orbit (away from
perigee) thus not impacting the useful measurement time (see
Ref.~\cite{Schubert2013} and the STE-QUEST Yellow
Book~\cite{YellowBook}).

The final uncertainty of the UFF/WEP test can be compared to present
and upcoming tests, by considering the corresponding E\"otv\"os
ratios. Figure~\ref{fig:testWEP} presents such a comparison for
different tests in the Earth field (except LLR which is in the Sun
field).

When examining Fig.~\ref{fig:testWEP}, one should bear in mind that
the compared experiments all use different test masses, and that thus
a direct comparison can be misleading, as discussed in
Sec.~\ref{sec:quantum} and~\ref{sec:diffEEP}. The only experiment that
the STE-QUEST UFF/WEP test can be compared to directly is the ground
measurement of the differential free fall of the two \text{Rb}
isotopes~\cite{Fray2004, Bonnin2013} with respect to which STE-QUEST
represents an impressive improvement by eight orders of magnitude. The
same is true when comparing to other purely quantum matter wave
experiments~\cite{Schlippert2014, Tarallo2014}.  Even when comparing
to macroscopic tests, with best present ground tests from the
E\"ot-Wash group~\cite{Schlamminger2008} or LLR~\cite{Williams2004},
both at the $2 \times 10^{-13}$ level, STE-QUEST still represents an
improvement by two orders of magnitude. However, it is important to
stress here that STE-QUEST measurement is truly quantum in nature (see
Sec.~\ref{sec:quantum}), in particular:
\begin{enumerate}
\item The observable is the phase difference of interfering matter
  waves in a coherent superposition;
\item The coherent superposition is well separated spatially by
  $\sim 10\,\text{cm}$, more than two orders of magnitude larger
  than the size of the individual wave packet parts;
\item The atoms are condensed to a quantum degenerate state (Bose
  Einstein Condensate);
\item The coherence length of the atoms is of the order of a
  micrometer, many orders of magnitude larger than the de Broglie
  wavelength of the macroscopic test masses ($10^{-27}\,\text{m}$ or
  less).
\end{enumerate}
Ground~\cite{Dimopoulos2007}, parabolic flights~\cite{Geiger2011},
drop tower~\cite{Muntinga2013} and sounding rocket tests using
coherent matter waves are also likely to improve within the STE-QUEST
time frame. However, STE-QUEST does not suffer from inherent limits of
the ground laboratory environment (short free-fall times, gravity
gradients, perturbed laboratory environment, \textit{etc.}), which
will ultimately limit tests on ground. Moreover, unique advantages of
a satellite operation (see Sec.~2.5 in the Yellow
Book~\cite{YellowBook}) are the chief assets towards the performance
announced. This is somewhat akin to classical tests where the next
significant improvement is expected from going into space with the
$\mu$-SCOPE mission.

Finally, we note that so far no analysis in the field of other sources
(\textit{e.g.} galactic dark matter~\cite{Schlamminger2008}) has been carried
out for STE-QUEST. This might lead to further interesting limits and
experimental possibilities, \textit{e.g.} by considering parts of the orbit
that are not useful for UFF/WEP in the Earth field, but are useful in
the field of more distant bodies. Such analysis and corresponding
optimization of the measurement scenario will be carried out as the
mission progresses and will further enhance the scientific discovery
potential of STE-QUEST.

\subsection{STE-QUEST Test of Local Position Invariance}
\label{sec:STElpi}

\begin{figure}[t]
\begin{center}
\begin{tabular}{c}
\includegraphics[width=13cm]{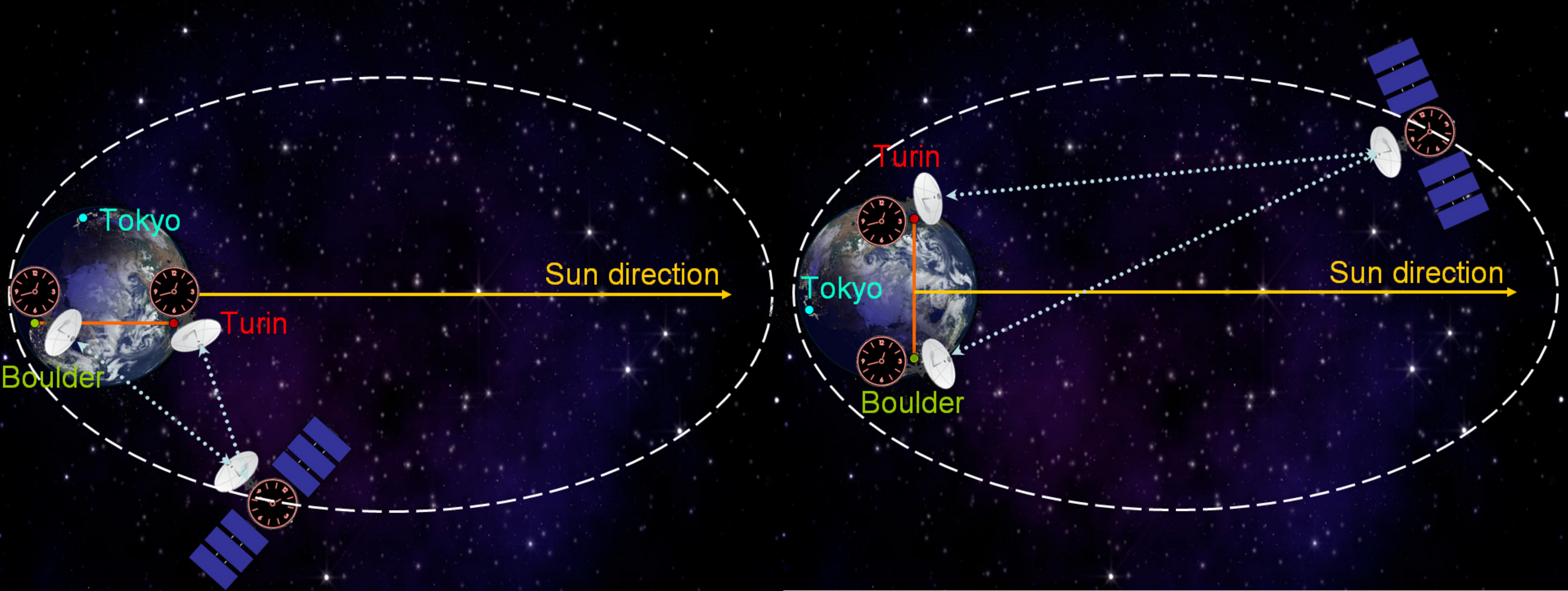}
	\end{tabular}
        \caption{Common-view comparison between Turin and Boulder for
          the test of LPI in the field of the Sun. The two panels show
          the different locations of the clocks in the field of the
          Sun as the Earth rotates, and their common-view comparison
          by STE-QUEST.}\label{fig:compareBT}
\end{center}
\end{figure}

In the baseline configuration (without clock on-board), STE-QUEST will
be able to compare distant ground clocks using the microwave link
(MWL) in common-view mode. In the common-view technique, two ground
clocks are simultaneously compared. The difference of simultaneous
measurements provides then a direct comparison of the two clocks on
the ground. This measurement does not require a high-performance
frequency reference on-board the STE-QUEST spacecraft. Indeed, the
noise of the space clock, which appears as common mode in the two
simultaneous link measurements, is rejected to high degree when the
difference of the two space-to-ground comparisons is
evaluated. According to the STE-QUEST reference orbit, common-view
contacts between USA and Europe, Europe and Japan, Japan and USA have
uninterrupted durations longer than 10 hours with each of them
repeated every two days. The concept of the LPI test in the
gravitational field of the Sun is shown in
Fig.~\ref{fig:compareBT}. In this example the frequency ratio
$\nu_\text{T}/\nu_\text{B}$ between two ground clocks in Turin and
Boulder is measured.

In the framework discussed in Sec.~\ref{sec:diffEEP}, we can consider
a generalization in which the Sun acts as the source of the anomalous
gravitational coupling. The measured frequency ratio of the two clocks
can be written as
\begin{equation}\label{nuBT}
  \frac{\nu_\text{T}}{\nu_\text{B}} = 1
  -\frac{1}{c^2}\left[U^\odot_\text{B}-U^\odot_\text{T} +
    \frac{v_\text{B}^2-v_\text{T}^2}{2} + \alpha^\odot_\text{B} U^\odot_\text{B} -
    \alpha^\odot_\text{T} U^\odot_\text{T} \right] + \Delta\,,
\end{equation}
where $U^\odot_\text{B}$ and $U^\odot_\text{T}$ are the solar
Newtonian gravitational potentials at the locations of the ground
clocks and $v_\text{B}$ and $v_\text{T}$ are the corresponding
velocities in a solar-system barycentric reference frame. The LPI
violating parameters $\alpha^\odot_\text{B}$ and
$\alpha^\odot_\text{T}$ depend on the type of transition used in the
respective clocks and possibly on the source of the gravitational
field (here the Sun); $\Delta$ represents all corrections due to the
other solar system bodies (including the Earth) assumed to behave
normally, as well as higher-order correction terms.

An essential point to note is that, in the absence of an LPI violation
($\alpha^\odot_\text{B}=\alpha^\odot_\text{T}=0$), the leading part in
Eq.~\eqref{nuBT} is equal to zero (up to small tidal correction terms
in $\Delta$ and constant terms from the Earth field). This is a direct
consequence of the EEP, as the Earth is freely falling in the Sun
field~\cite{Hoffman1961}. The LPI test in the Sun field is thus a
null test, verifying whether the measured frequency ratio is equal to
the expected value, \textit{i.e.} $1+\Delta$ in this example.

In general, the types of clocks used at the different ground stations
may be of different type so
$\alpha^\odot_\text{B}\not=\alpha^\odot_\text{T}$. In the following,
we will assume for simplicity clocks of the same type which simplifies
the LPI violating term in~\eqref{nuBT} to
$\alpha^\odot(U^\odot_\text{B}-U^\odot_\text{T})$, with the aim of the
experiment being the measurement of $\alpha^\odot$. More precisely the
experiment will measure the time evolution of the ratio
$\nu_\text{T}/\nu_\text{B}$, which again should be one in GR (up to
correction terms), but will evolve in time if the LPI violating
parameter is non-vanishing because of the time evolution of
$U^\odot_\text{B}-U^\odot_\text{T}$, mainly related to the rotation of
the Earth. The time evolution of
$(U^\odot_\text{B}-U^\odot_\text{T})/c^2$ will be predominantly
periodic with a daily period and peak-to-peak amplitude of about $1
\times 10^{-12}$.

Then, the determination of the LPI parameters boils down to a search
for a periodic signal with known frequency and phase in the clock
comparison data. As detailed in Ref.~\cite{YellowBook}, in the
baseline configuration the measurement uncertainties of the MWL and
the ground clocks should allow a detection of any non-zero value of
the LPI violating parameter $\alpha$ in the Sun field that exceeds $2
\times 10^{-6}$ after four years of integration. In the case that the
optional optical link is included in the payload, that goal can be
reached in 72 days of integration with the ultimate performance of $5
\times 10^{-7}$ reached in 4 years. Note however, that these results
are based on only frequency measurements without making use of the
phase cycle continuity provided by the STE-QUEST MWL. When phase cycle
continuity is maintained by the link, the measurement duration is not
affected by the dead-time between one common-view comparison and the
next, resulting in a reduction of the integration time needed to reach
the ultimate accuracy. Such a data analysis approach is presently
being implemented in the numerical simulations~\cite{YellowBook}.

The procedure for the LPI test in the Moon field is identical to the
Sun field test described above. The difference is that the frequency
and phase of the signal that one searches for are different and that
the sensitivity is decreased by a factor $\sim 175$, see
Eq.~\eqref{sunmoon}.

In the case where the onboard clock option of STE-QUEST is realized,
it will be possible to perform also an LPI test in the field of the
Earth. Given that this is only an option, we will discuss the test and
the results that can be achieved only briefly. Some more details can
be found in Ref.~\cite{YellowBook}. In this case the MWL (or optical)
link is used to compare the onboard clock to ground clocks. In the
formalism of Sec.~\ref{sec:diffEEP}, the frequency ratio can be written as
\begin{equation}\label{nuBSTE}
\frac{\nu_\text{STE}}{\nu_\text{B}} = 1
-\frac{1}{c^2}\left[U^\oplus_\text{B}-U^\oplus_\text{STE} +
  \frac{v_\text{B}^2-v_\text{STE}^2}{2} + \alpha^\oplus_\text{B} U^\oplus_\text{B}
  - \alpha^\oplus_\text{STE} U^\oplus_\text{STE} \right] + \Delta'\,,
\end{equation}
where $U^\oplus_\text{B}$ and $U^\oplus_\text{STE}$ are the Earth's
Newtonian gravitational potentials at the locations of the ground
clock and the onboard clock, and $v_\text{B}$ and $v_\text{STE}$ are
the corresponding velocities in a geocentric reference frame. The LPI
violating parameters $\alpha^\oplus_\text{B}$ and
$\alpha^\oplus_\text{STE}$ depend on the type of transition used in
the respective clocks. As in Eq.~\eqref{nuBT}, $\Delta'$ represents
all corrections due to the other solar system bodies, as well as
higher-order or numerically smaller correction terms (see
Ref.~\cite{Blanchet2001} for more details on relativistic time and
frequency transfer).

The main difference with respect to the Sun LPI test above is that the
ground clocks are not freely falling in the field of the Earth, so
even in the absence of an LPI violation the frequency ratio is not one
and is varying in time with the eccentric orbit of STE-QUEST. The test
then compares the theoretically calculated frequency ratio (from the
knowledge of the STE-QUEST orbit and the ground station locations) to
the actually measured one. This leads to two methods for the
measurement, one based on the accuracy of the clocks (so-called DC
measurement) that searches for an offset with respect to the expected
value, and one based on the periodic variation due to the orbit
eccentricity (so-called AC measurement) that searches for the time
varying signature and thus relies on the clock stability. The former
is carried out mainly when the satellite is at apogee, when the LPI
violating term in~\eqref{nuBSTE} is largest; the latter uses
measurements over the full orbit. As detailed in
Ref.~\cite{YellowBook}, simulations taking into account the MWL and
clock noise and accuracy show that with both methods an uncertainty of
$2 \times 10^{-7}$ on the LPI violating parameter $\alpha^\oplus$ can
be reached after 4 days (DC measurement) and 840 days (AC measurement)
of integration. In the case of the DC measurement, the limit is
imposed by the clock accuracy rather than the measurement duration. In
the case of the AC measurement the uncertainty can be decreased to
$1.5 \times 10^{-7}$ when integrating over the whole mission duration.
\begin{figure}[t]
\begin{center}
\begin{tabular}{c}
\includegraphics[width=16cm]{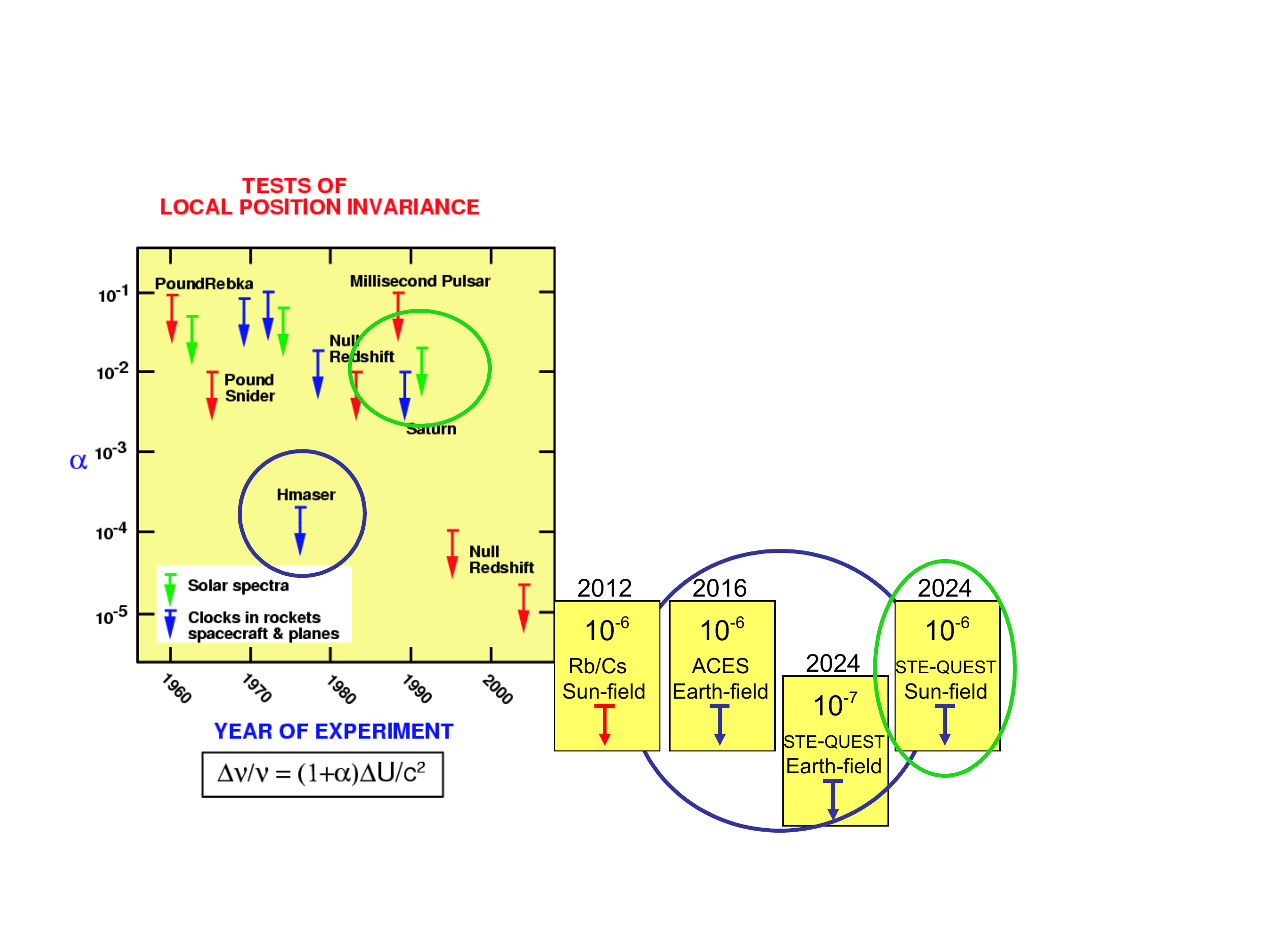}
	\end{tabular}
\caption{Present and upcoming limits on LPI violation adapted from
  Ref.~\cite{Will2006}. Red arrows (after 1990) represent ``null
  red-shift measurements'' that can only provide the difference
  $\alpha_i-\alpha_j$ for two different types of clocks $i$ and
  $j$. Inside the green circles are the best direct LPI tests in the
  Sun field providing directly $\alpha^\odot_i$ for the respective
  transition. Inside the blue circles are the best direct LPI tests in
  the Earth field for $\alpha^\oplus_i$ (optional in the case of
  STE-QUEST).}\label{fig:testLPI}
\end{center}
\end{figure}

The sensitivities of STE-QUEST estimated above can be compared to
present and upcoming LPI tests by looking directly at the limits on
the corresponding parameters (see Sec.~\ref{sec:diffEEP}). Such a
comparison is presented in Fig.~\ref{fig:testLPI}, adapted and updated
from Ref.~\cite{Will2006}.

Figure~\ref{fig:testLPI} shows a number of experiments, including null
tests and direct tests. The latter set limits directly on the
parameter $\alpha_i$ for the relevant transition, \textit{e.g.} the
\text{H}-maser experiment of 1979~\cite{Vessot1979} sets a limit on
$\alpha^\oplus_\text{H}$ for the hydrogen hyperfine transition in the
Earth gravitational field. The ``Null Red-shift'' experiments in
Fig.~\ref{fig:testLPI} consist of two co-located clocks of different
type in the same laboratory whose relative frequency is monitored as
the local gravitational potential varies in time. Thus one measures
$(\alpha_i-\alpha_j)U/c^2$ for two clocks of type $i$ and $j$ and sets
a limit on the difference $\alpha_i-\alpha_j$. The most precise such
test at present sets a limit of
$\alpha^\odot_\text{Rb}-\alpha^\odot_\text{Cs}=(0.11\pm 1.0)\times
10^{-6}$ for the $\text{Rb}$ \textit{vs.} $\text{Cs}$ hyperfine
transitions~\cite{Guena2012}, using the annual variation of the solar
potential in the laboratory due to the eccentricity of the Earth's
orbit. Depending on the underlying model, the difference
$\alpha_i-\alpha_j$ might be much smaller than the individual values,
especially when similar transitions are used (both hyperfine or both
electronic, \textit{i.e.}  optical), so direct tests that measure
$\alpha_i$ individually rather than differences $\alpha_i-\alpha_j$
are necessary and complementary to co-located tests, which is one of
the main drivers for experiments like ACES or STE-QUEST. In the
STE-QUEST LPI test, a non-zero signal will be observed no matter what
the actual values of $\alpha^\odot_\text{T}$ and
$\alpha^\odot_\text{B}$ in Eq.~\eqref{nuBT} are, provided at least one
of them is non-zero, because of the different temporal variations of
$U^\odot_\text{B}$ and $U^\odot_\text{T}$. This is not the case in
null-tests with co-located clocks, where one necessarily has $U_i=U_j$
and thus a signal can only be detected if $\alpha_i\not=\alpha_j$,
which is not the case for STE-QUEST.

Finally, all Sun LPI science objectives also apply to a test with the
Moon as the source mass. STE-QUEST will carry out a direct LPI test in
the Moon field using the same methods (and data) as the test in the
Sun field described above. Note that the two putative signals can be
easily de-correlated in the data due to the different frequency and
phase. The sensitivity of STE-QUEST to a possible violation of LPI
sourced by the Moon is then simply given by a reduction factor with
respect to the Sun effect of
\begin{equation}\label{sunmoon}
\left(\frac{M_\text{Sun}}{d_\text{Sun}^2}\right)/
\left(\frac{M_\text{Moon}}{d_\text{Moon}^2}\right) = 175\,.
\end{equation}
In the baseline configuration the measurement uncertainties of the MWL
and the ground clocks should allow a detection of any non-zero value
of the LPI violating parameter $\alpha$ sourced by the Moon that
exceeds $4 \times 10^{-4}$ after four years of integration. In the
case that the optional optical link is included in the payload, that
goal can be reached in 72 days of integration with the ultimate
performance of $9 \times 10^{-5}$ reached in 4 years. Like for the Sun
test, note that these results are based on only frequency measurements
without making use of the phase cycle continuity requested for the
MWL.

Clock tests as described above are sometimes interpreted as searches
for a space-time variation of fundamental constants, in particular
those of the Standard Model (fine structure constant, electron, proton
and quark masses, QCD mass scale, \textit{etc.}). Such an
interpretation is model-dependent (one assumes the validity of the
Standard Model of particle physics to describe atomic transitions) so
we do not use it here as our aim is to remain as general as
possible. In order to best constrain all possible variations of
constants the comparison of as many different transitions as possible
is essential. Comparisons of ground clocks based on different types of
transitions repeated during the STE-QUEST mission (see the Yellow
Book~\cite{YellowBook}) will provide a wealth of data to search for
temporal variations of fundamental constants, the fine structure
constant $\alpha$ and the electron-to-proton mass ratio $\mu$ in
particular. Different clock transitions have different dependency on
fundamental constants. Therefore, the results of crossed frequency
comparisons repeated in time provides a clear interpretation of any
observed drift over time and imposes unambiguous limits on time
variations of fundamental constants. Current best limits on time
variations of fundamental constants from laboratory experiments are
consistent with zero: the $\alpha$ drift was recently measured to
$\dot{\alpha}/\alpha = (5.8\pm 6.9)\times 10^{-17}\,\text{yr}^{-1}$ in
the dysprosium experiment~\cite{Leefer2013}; the $\alpha$ and $\mu$
drifts were determined to $\dot{\alpha}/\alpha = (-1.6\pm 2.3)\times
10^{-17}\,\text{yr}^{-1}$ and $\dot{\mu}/\mu = (1.9\pm 4.0)\times
10^{-16}\,\text{yr}^{-1}$ in the frequency comparison of a
$\text{Hg}^+$ and an $\text{Al}^+$ clock~\cite{Rosenband2008}. At the
same time, data obtained from astronomical observation of quasar
absorption spectra are providing complementary information exploring
completely different measurement systematics: $\Delta\alpha/\alpha =
(-0.57 \pm 0.11)\times 10^{-5}$~\cite{Murphy2004} and
$\vert\Delta\mu/\mu\vert < 1.8 \times 10^{-6}$ (95\% confidence
level)~\cite{Murphy2008}, at approximately half the Universe's current
age. Interestingly enough, even if their interpretation is still
controversial, these data seem to indicate a time variation of
$\alpha$. Additional and more precise measurements are clearly needed
to better understand and resolve the puzzle. These limits are expected
to improve by at least one order of magnitude thanks to STE-QUEST.

\subsection{STE-QUEST Tests of Lorentz Invariance and CPT Symmetry}
\label{sec:STEcpt}

Lorentz Invariance is the third sub-principle of the EEP as described
in Sec.~\ref{sec:facet}. Currently, there is a great deal of interest
in Lorentz Invariance and the combined charge conjugation, parity,
time reversal (CPT) symmetry --- and, in particular, the question of
whether these related symmetries are truly exact in Nature. Both the
Standard Model of particle physics and GR are precisely invariant
under (local) Lorentz and CPT symmetries, which makes these symmetries
particularly fundamental. Whilst Lorentz and CPT symmetries have been
discussed frequently in frameworks similar to the one introduced in
Sec.~\ref{sec:diffEEP}, a more general, broad, and complete framework
for tests of Lorentz Invariance and CPT symmetry has been developed
over the last decade: the Standard Model Extension
(SME)~\cite{Colladay1997, Colladay1998}. We will use this framework to
analyze Lorentz Invariance tests that will be carried out with
STE-QUEST. The first estimates presented here give a general idea of
the potential of STE-QUEST in this field.

Many candidate theories of quantum gravity suggest the possibility of
Lorentz and CPT symmetry breaking in certain regimes. For example, the
symmetries could be broken spontaneously, either in string
theory~\cite{Kostelecky1989, Kostelecky1991} or in quantum field
theories with fundamental tensor fields~\cite{Altschul2005,
  Altschul2010}. There could also be Lorentz-violating physics in loop
quantum gravity~\cite{Gambini1999, Alfaro2002} and non-commutative
geometry theories~\cite{Mocioiu2000, Carroll2001}; Lorentz violation
through spacetime-varying couplings~\cite{Kostelecky2003,
  Ferrero2009}; or breaking of Lorentz and CPT symmetries by quantum
anomalies in certain space-times with nontrivial
topologies~\cite{Klinkhamer2004}. Moreover, since CPT violation in a
well-behaved low-energy effective quantum theory automatically
requires Lorentz violation as well~\cite{Greenberg2002}, any
predictive theory that entails violations of CPT will also include
violations of Lorentz Invariance.

So far, there is no compelling evidence that Lorentz and CPT
symmetries are not actually exact in Nature. In fact, there have been
numerous experimental tests of these theories, using a very wide
variety of techniques. Recent experimental tests have included studies
of matter-antimatter asymmetries for trapped charged particles and
bound state systems, measurements of muon properties, analyses of the
behavior of spin-polarized matter, frequency standard comparisons,
Michelson \& Morley-type experiments with resonators, Doppler effect
measurements, measurements of neutral meson oscillations, polarization
measurements on the light from cosmological sources, high-energy
astrophysical tests, precision tests of gravity, and others (see
Ref.~\cite{Kostelecky2011a} for a compilation of the present
experimental constraints).

A general effective field theory that describes Lorentz violation for
elementary particles is the SME~\cite{Colladay1997, Colladay1998}. As
a quantum field theory, the SME contains all Lorentz-violating
operators that can be written down using Standard Model fields, along
with coefficients for Lorentz violation that parameterize the
Lorentz-violating effects. These coefficients vanish in a perfectly
Lorentz-invariant theory. It has also been expanded (as a classical
field theory) to give a systematic way of studying Lorentz-violating
and CPT-violating gravitational
interactions~\cite{Kostelecky2004}. The SME gravitational action
includes both space-time curvature effects and space-time torsion
phenomena; some of the torsion effects turn out to be equivalent, at
least locally, to spin-dependent Lorentz-violating operators in the
particle physics sectors of the SME~\cite{Kostelecky2008}. Although
many theories describing new physics suggest the possibility of
Lorentz violation, none of them are understood well enough to make
firm predictions. The greatest utility of the SME is the theory's
generality. The SME provides a framework for placing constraints on
Lorentz and CPT-violating effects, without worrying about the
underlying mechanism by which the symmetry violation arises.

In the presence of Lorentz invariance violation, experimental results
will depend on the orientation of the apparatus (for violations of
spatial isotropy) and on the velocity of the apparatus (for violations
of Lorentz boost invariance). For Earthbound experiments the changes
of orientation and velocity are limited to the Earth rotation and
orbital motion or to slow modulations imposed in the laboratory
(\textit{e.g.}  turntables). For a satellite experiment, there are new
forms of motion, and this enhances the sensitivity to Lorentz
violation. The highly eccentric and time-varying orbit of the
STE-QUEST satellite will be extremely advantageous for several
reasons. Tests of Lorentz boost symmetry require comparisons of data
collected in different Lorentz frames. It is necessary to physically
boost the experiment into different frames and compare the results
observed under the different conditions. The sensitivity to boost
invariance violations is then determined by the velocity differences
$\bm{v}$ between different observation frames. The direction of
$\bm{v}$ determines the specific linear combination of violation
coefficients that can be constrained by a single
comparison. Simultaneously, the speed determines the strength of the
constraint; for nonrelativistic relative speeds, $v/c\ll 1$ is a
direct suppression factor. For these reasons, it is advantageous to
sample as many frames, moving as rapidly in relation to one-another,
as possible.

When coupled to gravity, one finds that WEP/UFF tests can provide the
best available sensitivity to certain types of Lorentz violation in
the SME~\cite{Kostelecky2011b}. In fact, several Lorentz-violating
possibilities can only be tested using such precision gravitational
experiments~\cite{Kostelecky2009a}. Hence, the impressive WEP tests of
the STE-QUEST mission would provide the best sensitivities to date on
an additional set of coefficients for Lorentz violation.

Effective WEP violation in the SME originates from its generality in
allowing the possibility of coefficients for Lorentz violation that
differ among fermions of different flavors. That is, the degree of
Lorentz violation may differ from protons, to neutrons, to electrons,
for example. When gravitational couplings are considered for fermions,
this species dependence leads to a differing gravitational
response. Since the effect is due to Lorentz violation, variation in
the size of the effective WEP violation with the orientation and boost
direction of the experiment typically results, as do modifications in
the direction of the gravitational acceleration. Thus, WEP tests such
as those on the STE-QUEST mission typically have the ability to
distinguish a signal due to Lorentz violation from other sources of
WEP violation \textit{via} the dependence of the signal on orientation
and velocity as well as the unique direction dependence of the
acceleration.

The proposed WEP/UFF experiment is of the class that was analyzed
extensively by Kosteleck\'y~\cite{Kostelecky2011b}. Explicit
predictions obtained for experiments on Earth extend naturally to
STE-QUEST through replacement of appropriate boost and gravitational
factors. Performing such an experiment in space provides the benefits
of variable boost orientations and longer free-fall times, but there
is no fundamental change in the existing analysis. The results of that
analysis along with the WEP sensitivity goals of STE-QUEST imply that
sensitivities ranging from the $10^{-11}$ to $10^{-7}$ levels per
measurement cycle will be possible for up to 8 combinations of SME
coefficients. After incorporating data from the large number of orbits
throughout the mission, constraints ranging from $10^{-14}$ to
$10^{-10}$ levels are expected. These sensitivities would provide
improvements of up to 5 orders of magnitude over existing constraints.

Another way in which the STE-QUEST mission as currently proposed could
attain sensitivity to Lorentz violation is through red-shift
tests. Coefficients for Lorentz violation, which couple to
gravitational fields in the SME lead to modified space-time
curvature~\cite{Kostelecky2011b, Bailey2006} as well as additional
modifications to clock frequencies, and specific predictions for
red-shift experiments have been made~\cite{Kostelecky2011b,
  Bailey2009}. These predictions include that of a variation in the
red-shift signal as the clock explores the gravitational potential
that is qualitatively different from the conventional red-shift
signal~\cite{Bailey2009}. This effect arises due to the impact of
rotation-invariance violation on the gravitational field. While the
sensitivity to the relevant SME coefficients available \textit{via}
red-shift tests will not exceed the maximum reach currently available
\textit{via} other types of experiments, such tests are still
interesting from a SME perspective for two reasons. The sensitivities
to Lorentz violation achieved in a given experiment often constrain or
measure a large combination of coefficients from the theoretical
framework, hence additional tests can provide the necessary
information to disentangle these combinations. Secondly, the present
analysis of such experiments considers implications of the minimal
gravitationally coupled SME only. Higher dimension
operators~\cite{Kostelecky2009b, Kostelecky2012} for which specific
predictions for experiments of this type have not yet been made may
result in additional effects that can be measured in this way.

Additional tests of Lorentz Invariance are possible if the optional
onboard clock is flown. The dependences of the atomic clock
frequencies on the minimal SME parameters is already
known~\cite{Kostelecky1999a, Bluhm2003}. These dependences were
determined using the effective Hamiltonian that may be derived from
the SME Lagrangian~\cite{Kostelecky1999b}. The algorithm for
calculating the frequency shifts is quite general and can be applied
to virtually any atomic clock transition. In the context of the
Schmidt nuclear model~\cite{Schmidt1937}, all the angular momentum $I$
of odd-even nucleus is carried by a single unpaired nucleon. The
principal sensitivities (given by the Schmidt model and the atomic
shell structure) are to Lorentz violation coefficients in the proton
sector, with secondary sensitivities in the electron sector. Searching
for modulations in the transition frequency with the characteristic
satellite orbital frequency will make it possible to place constraints
on up to 25 coefficients in the proton sector, with sensitivity levels
ranging from $10^{-21}$ down to $10^{-28}$. A further 18
electron-sector coefficients may be constrained with potentially
$10^{-19}$ to $10^{-27}$ level sensitivities. For most of these
coefficients, these are unprecedented levels of sensitivity. There are
also dependences on additional SME coefficients, which are not
captured by the Schmidt model. These include dependences on neutron
coefficients and dependences that exist because of the relatively
rapid movement of the nucleons inside an atomic
nucleus~\cite{Altschul2009}. The extremely sensitive data provided by
STE-QUEST would allow extracting these additional dependences.

In conclusion of this section, STE-QUEST offers the possibility to
explore a large parameter space of the SME and to thereby constrain,
or uncover, violations of Lorentz and CPT symmetry. In particular,
coefficients in the proton and electron sector will be constrained
from the clock measurements, while the
${}^{85}\text{Rb}$--${}^{87}\text{Rb}$ atom interferometer will
provide new constraints in the gravitational sector, with expected
improvements of up to five orders of magnitude on present limits.

\section{Conclusions}
\label{sec:conclusion}

\begin{table*}[t]
\begin{center}
\begin{tabular}{|p{5.4cm}||p{11cm}|} 
  \hline \centerline{\textbf{Science Investigation}} & 
  \centerline{\textbf{Measurement
      Requirement}} \\ \hline \multicolumn{2}{|l|}{\textbf{Clock
      Comparisons and International Atomic Time Scales}} \\ \hline
  \textit{Common-view comparisons of ground clocks} & Common-view
  comparison of ground clocks at the $1\times 10^{-18}$ fractional
  frequency uncertainty level after a few days of integration time with
  the STE-QUEST microwave link and a few hours by using the optical
  link.  \\ \hline \textit{Space-to-ground time transfer} &
  Space-to-ground time transfer with accuracy better than
  $50\,\text{ps}$. \\ \hline Synchronization of ground clocks &
  Synchronization of clocks on ground to better than
  $50\,\text{ps}$. \\ \hline \textit{Atomic time scales} & Contribution
  to the generation of atomic time scales to fractional frequency
  inaccuracy lower than $1\times 10^{-16}$. \\ \hline \textit{GNSS
    clocks and time scales (optional)\footnote{This scientific
      investigation can be performed only if the STE-QUEST payload is
      equipped with a high-stability and high-accuracy atomic clock.}} &
  Monitoring of the stability of on-board GPS, GALILEO, and GLONASS
  clocks. \\ \hline \multicolumn{2}{|l|}{\textbf{Geodesy}}\\ \hline
  \textit{On-site differential geopotential measurements} & Differential
  geopotential measurements between two points on the Earth's surface
  with resolution in the gravitational potential $U$ at the level of
  $0.15\,\text{m}^2/\text{s}^2$ (equivalent to $1.5\,\text{cm}$ on the
  geoid height difference). \\ \hline
  \multicolumn{2}{|l|}{\textbf{Reference Frames}}\\ \hline \textit{Earth
    terrestrial and celestial reference frame} & Realization and
  unification of the terrestrial and the celestial reference frame of
  the Earth. \\ \hline
\end{tabular}
\caption{Science investigations \textit{vs.} measurement requirements
  for topics other than fundamental physics that shall be investigated
  by STE-QUEST.}\label{tab:othertopics}
\end{center}
\end{table*}
We have presented the fundamental physics science objectives of
STE-QUEST, which are centered on tests of the three different aspects
of the Einstein Equivalence Principle (EEP): the Weak Equivalence
Principle (WEP) or Universality of Free Fall, Local Position
Invariance (LPI) or Universality of Clock Rates, and Local Lorentz
Invariance (LLI) coupled to CPT symmetry. One of the unique strengths
of STE-QUEST is that it will test all three aspects of the EEP, using
a combination of measurements in space and on the ground (relative
acceleration of different atomic isotopes, comparison of distant
clocks). Although the three sub-principles are connected by Schiff's
conjecture, the actual quantitative merit of the different experiments
is model-dependent (see Sec.~\ref{sec:diffEEP}). As a consequence, it
is not known \textit{a priori} which test (WEP, LPI, or LLI) is more
likely to first detect a violation and the most reasonable approach is
to pursue tests of the three sub-principles with equal vigor. This is
the baseline of STE-QUEST, which carries out, and improves on,
state-of-the-art tests of all three sub-principles.

Another unique feature of STE-QUEST is its capability to carry out the
WEP test using quantum matter waves in superpositions that have no
classical counterpart. Although, we know of no viable model that
predicts an EEP violation specific to such quantum matter waves, one
should be prepared for surprises as there are numerous open questions
at the interface between gravitation and quantum mechanics reviewed in
Sec.~\ref{sec:quantum}. Those issues provide good reason for exploring
the foundations of general relativity, \textit{i.e.} the EEP, with as
diverse objects as possible, like antimatter or quantum degenerate
gases and superposition states.

Finally let us mention that although the primary science objectives of
STE-QUEST are in fundamental physics, the mission will also provide a
wealth of legacy science for other fields like time/frequency
metrology, reference frames and geodesy. These are summarized in
Table~\ref{tab:othertopics}, with more details available in the
STE-QUEST Yellow Book~\cite{YellowBook}.

To conclude, all of this gives a unique science case for STE-QUEST,
making it the first space mission to carry out such a complete test of
the foundations of gravitation theory using quantum sensors, and
providing additionally some legacy science for other applications. If
selected in an upcoming call, it will follow in the wake of precursor
missions like ACES and $\mu$-SCOPE (to be launched by ESA and CNES in
2016) and firmly establish Europe's lead in fundamental physics as a
space science discipline.

\begin{acknowledgments}
  The authors would like to acknowledge support from ESA and National
  space agencies (CH, D, ES, F, GR, I, S, UK). Significant support was
  provided by numerous scientists strongly involved in the elaboration
  of the science case, in the simulation activities, in the payload
  and instruments studies.
\end{acknowledgments}

\appendix
%


\end{document}